\documentclass[aps,prl,twocolumn,superscriptaddress,nobibnotes,showpacs]{revtex4-1}
\usepackage{gensymb}
\usepackage{graphicx}
\usepackage{xcolor}
\usepackage{ulem}
\usepackage{bm,hyperref}
\usepackage{amssymb}

\begin{document}

\title{\textbf{Evidence of Fermi Surface Reconstruction at the Metamagnetic Transition of the Strongly Correlated Superconductor UTe$_2$}}

\author{Q.~Niu}
\affiliation{Univ. Grenoble Alpes, CEA, IRIG, PHELIQS, F-38000 Grenoble, France}
\author{G.~Knebel}
\affiliation{Univ. Grenoble Alpes, CEA, IRIG, PHELIQS, F-38000 Grenoble, France}
\author{D.~Braithwaite }
\affiliation{Univ. Grenoble Alpes, CEA, IRIG, PHELIQS, F-38000 Grenoble, France}
\author{D.~Aoki }
\affiliation{Univ. Grenoble Alpes, CEA, IRIG, PHELIQS, F-38000 Grenoble, France}
\affiliation{Institute for Materials Research, Tohoku University, Oarai, Ibaraki, 311-1313, Japan}
\author{G.~Lapertot}
\affiliation{Univ. Grenoble Alpes, CEA, IRIG, PHELIQS, F-38000 Grenoble, France}
\author{M.~Vali\v{s}ka}
\affiliation{Univ. Grenoble Alpes, CEA, IRIG, PHELIQS, F-38000 Grenoble, France}
\author{G.~Seyfarth}
\affiliation{Univ. Grenoble Alpes, EMFL, CNRS, Laboratoire National des Champs Magn\'etiques Intenses (LNCMI), 38042 Grenoble, France}
\author{W.~Knafo}
\affiliation{Laboratoire National des Champs Magnétiques Intenses, UPR 3228, CNRS-UPS-INSA-UGA,143 Avenue de Rangueil, 31400 Toulouse, France}
\author{T. Helm}
\affiliation{Max Planck Institute for Chemical Physics of Solids, 01187 Dresden, Germany}
\affiliation{Dresden High Magnetic Field Laboratory (HLD-EMFL), Helmholtz-Zentrum Dresden-Rossendorf, 01328 Dresden, Germany}
\author{J-P.~Brison}
\affiliation{Univ. Grenoble Alpes, CEA, IRIG, PHELIQS, F-38000 Grenoble, France}
\author{J.~Flouquet}
\affiliation{Univ. Grenoble Alpes, CEA, IRIG, PHELIQS, F-38000 Grenoble, France}
\author{A.~Pourret}
\email[E-mail me at: ]{alexandre.pourret@cea.fr}
\affiliation{Univ. Grenoble Alpes, CEA, IRIG, PHELIQS, F-38000 Grenoble, France}
\date{\today }

\begin{abstract}

Thermoelectric power ($S$) and Hall effect ($R_\mathrm{H}$) measurements on the paramagnetic superconductor UTe$_2$ with magnetic field  applied along the hard magnetization $b$-axis are reported. The first order nature of the metamagnetic transition at $H_\mathrm{m}=H^b_\mathrm{c2}=35$~T leads to drastic consequences on $S$ and $R_\mathrm{H}$. In contrast to the field dependence of the specific heat in the normal state through $H_\mathrm{m}$, $S(H)$ is not symmetric with respect to $H_\mathrm{m}$. This implies a strong interplay between ferromagnetic (FM) fluctuations and a Fermi-surface reconstruction at $H_\mathrm{m}$. $R_\mathrm{H}$ is very well described by incoherent skew scattering above the coherence temperature $T_\mathrm{m}$ corresponding roughly to the temperature of the maximum in the susceptibility $T_{\chi_\mathrm{max}}$ and coherent skew scattering at lower temperatures.
The discontinuous field dependence of  both, $S(H)$ and the ordinary Hall coefficient $R_0$, at $H_\mathrm{m}$ and at low temperature, provides evidence of a change in the band structure  at the Fermi level.


\end{abstract}


\maketitle

The recent discovery of unconventional superconductivity (SC) in the uranium chalcogenide paramagnet UTe$_2$ with a superconducting transition temperature $T_\mathrm{SC}\sim1.6$~K \cite{Ran2018,Aoki2019, Knebel2019} opens new perspectives on superconducting topological properties including emergent Majorana quasiparticles at the verge of magnetic and electronic instability. Transport and thermodynamic measurements demonstrated that correlations play an important role in this system requiring theoretical treatment beyond LDA approach \cite{Niu2019,Aoki2019,Ishizuka2019, Fujimori2019, Xu2019, Miao2019}. The closeness of  UTe$_2$ to a ferromagnetic quantum criticality \cite{Tokunaga2019} induces astonishing superconducting properties. Indeed when the magnetic field is applied along the hard $b$-axis at low temperature, superconductivity survives up to an extremely high field, $H_{c2}=35$~T, where it is destroyed abruptly by the occurrence of a  huge metamagnetic transition  (MMT) at $H_\mathrm{m}=H_\mathrm{c2}$  \cite{Knebel2019,Ran2019}. The unconventionnal  superconducting state in this system, i.e. spin-triplet Cooper pairing, has been identified by a small decrease in the NMR Knight shift \cite{Nakamine2019} and the large $H_{c2}$ exceeding the Pauli-limiting field  \cite{Ran2018,Aoki2019, Knebel2019}. Furthermore, re-entrant superconductivity (RSC) arises above $H_\mathrm{m}$ when magnetic field is tilted 30$^{\circ}$ away from the $b$ axis towards the $c$ axis \cite{Ran2019}. 
The MMT occurring at $H_\mathrm{m}$ with a jump in the magnetization of $0.6~\mu_B$, when the systems enters the polarised paramagnetic state (PPM) \cite{Miyake2019, Ran2019, Imajo2019}, is in agreement with a characteristic energy scale given by the temperature of the maximum in the susceptibility $T_{ \chi_\mathrm{max}}\approx 35$~K \cite{Knafo2019} and the maximum of the Hall effect ($R_\mathrm{H}$) \cite{Niu2019}.  Furthermore, fluctuations are strongly enhanced through $H_\mathrm{m}$ despite the first order nature of the MMT below the critical end point (CEP) at $T_\mathrm{CEP} \approx 7$~K \cite{Knafo2019, Miyake2019}. 

By some aspects, UTe$_2$ has properties similar to that found in the unconventional ferromagnetic superconductor URhGe \cite{Aoki2001}. It shows similar field enhancement of the Sommerfield coefficient $\gamma$ (linear $T$ term of the specific heat) associated to reentrant superconductivity (RSC) when approaching $H_\mathrm{m}$ ($H_\mathrm{m}=H_\mathrm{r} \approx 11.75$~T in URhGe)\cite{Miyake2008a,Hardy2011a,Wu2017}. In URhGe, the MMT is connected to a Fermi-surface instability \cite{Gourgout2016, Aoki2014} which may drive the SC \cite{Yelland2011,Sherkunov2018}. In UTe$_2$ as well as in URhGe, the MMT occurs for field along the hard magnetization axis. In both systems the MMT is strongly connected to the field enhancement of SC. A major difference is that URhGe is ferromagnetic with a Curie temperature $T_\mathrm{C}=9.5$~K at $H=0$ while UTe$_2$ remains paramagnetic (PM) at least down to 20~mK \cite{Sundar2019,Paulsen2020}.

A key question is the respective roles of ferromagnetic fluctuations and Fermi-surface instabilities at the MMT where the SC is abruptly suppressed. Indeed, the large step-like increase of the residual term of the resistivity, $\rho_0$, at the MMT suggests that in addition to magnetic fluctuations, a change in the carrier density may occur at the MMT for $H \parallel b$ \cite{Knafo2019}. 
For this purpose, we investigated the temperature and magnetic field dependences of the Seebeck coefficient ($S$) up to 36~T and the Hall resistance ($R_\mathrm{H}$) up to 68~T of $\mathrm{UTe}_2$ for $H \parallel b$. RSC is observed in both $S$ and $R_\mathrm{H}$ close to $H_\mathrm{m}$ around 1~K consistent with resistivity results \cite{Knebel2019}. The drastic changes in $S$ and in the ordinary Hall effect ($R_0$) at $H_\mathrm{m}$ point to a Fermi-surface reconstruction, contrasting with the rather symmetric behaviour of the $\gamma$ term \cite{Miyake2019} and of the $A$ coefficient ($T^2$ term of the resistivity) through $H_\mathrm{m}$ \cite{Knafo2019}.

\begin{figure}
\includegraphics[width=0.5\textwidth]{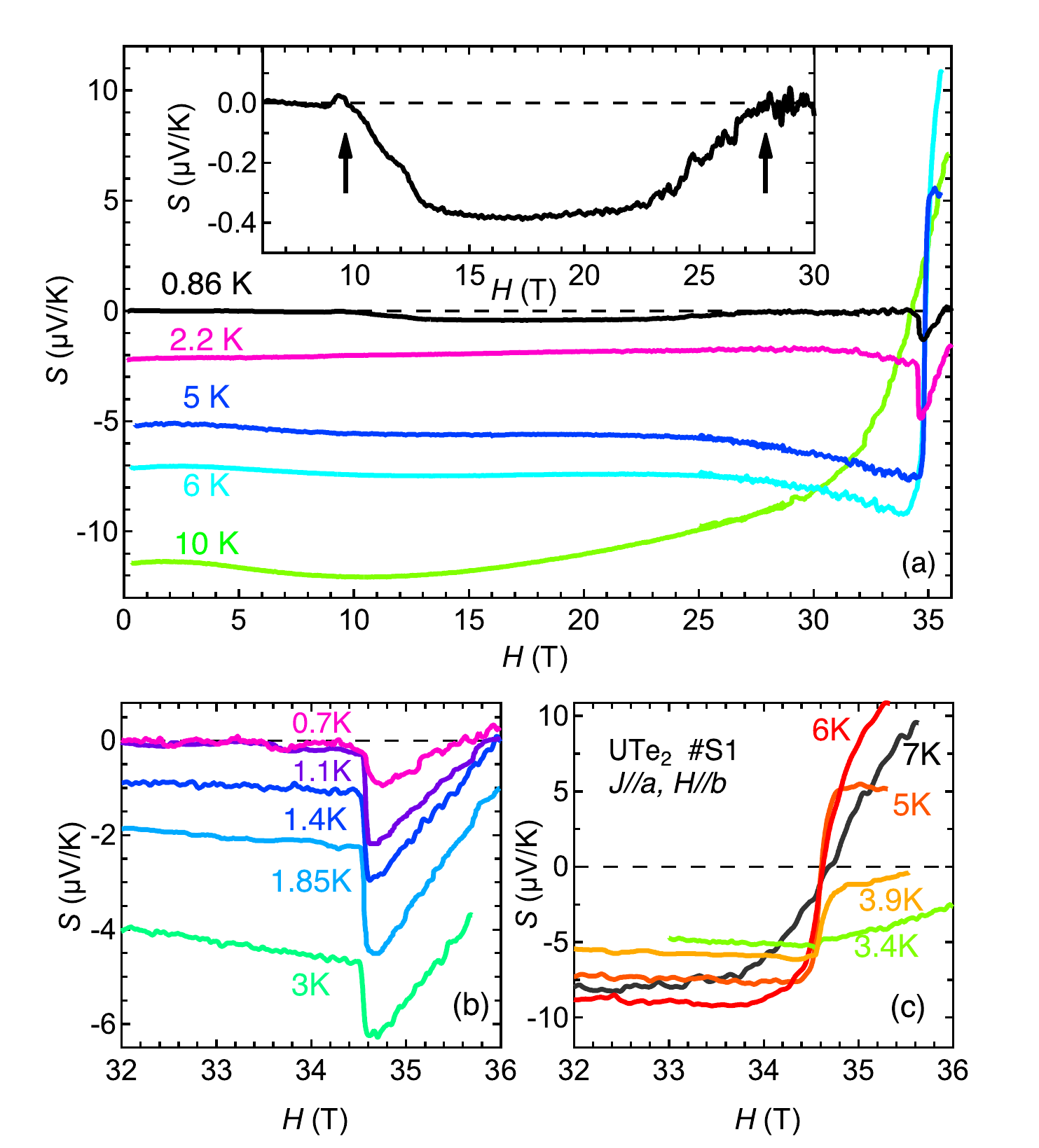}
\caption{(color online) Field dependence of $S$ in UTe$_2$ for $H \parallel b$ between 0 and 36~T (a) and between 32 and 36~T (b-c). $S(H)$ shows a clear anomaly at $H_\mathrm{m}$= 34.6~T, which broadens with increasing temperature. This anomaly changes from a negative to a positive jump above 3.5~K. The inset in (a) shows the RSC in $S(H)$ at 0.86~K in the field range from 6~T to  30~T.}
\label{Fig1}
\end{figure}

Single crystals of UTe$_2$ were grown by chemical vapor transport with iodine as transport agent. The orientation of the crystals has been verified by Laue diffraction. We performed the $S$,  $\rho$ and $R_\mathrm{H}$ measurements on three samples labelled S1, S2 and S3 with a residual resistivity ratio (RRR=$\frac{\rho(300~K)}{\rho(1.5~K)}$) of 30, 30 and 22 respectively. The samples were prepared for experiments with heat or electric current along the $a$-axis and the magnetic field along the $b$-axis. $S$ and $R_\mathrm{H}$ have been measured on sample S1 using a standard ``one heater-two thermometers" setup and $\rho_{xx}$ and $R_\mathrm{H}$ on sample S2 and S3 with a standard 6-probe method. The temperature and field dependences of different transport properties have been measured at LNCMI Grenoble using a $^3$He cryostat up to 36~T and on sample S3 at LNCMI Toulouse in pulsed field up to 68~T and tempearature down to 1.5~K.

\begin{figure}
\includegraphics[width=0.5\textwidth]{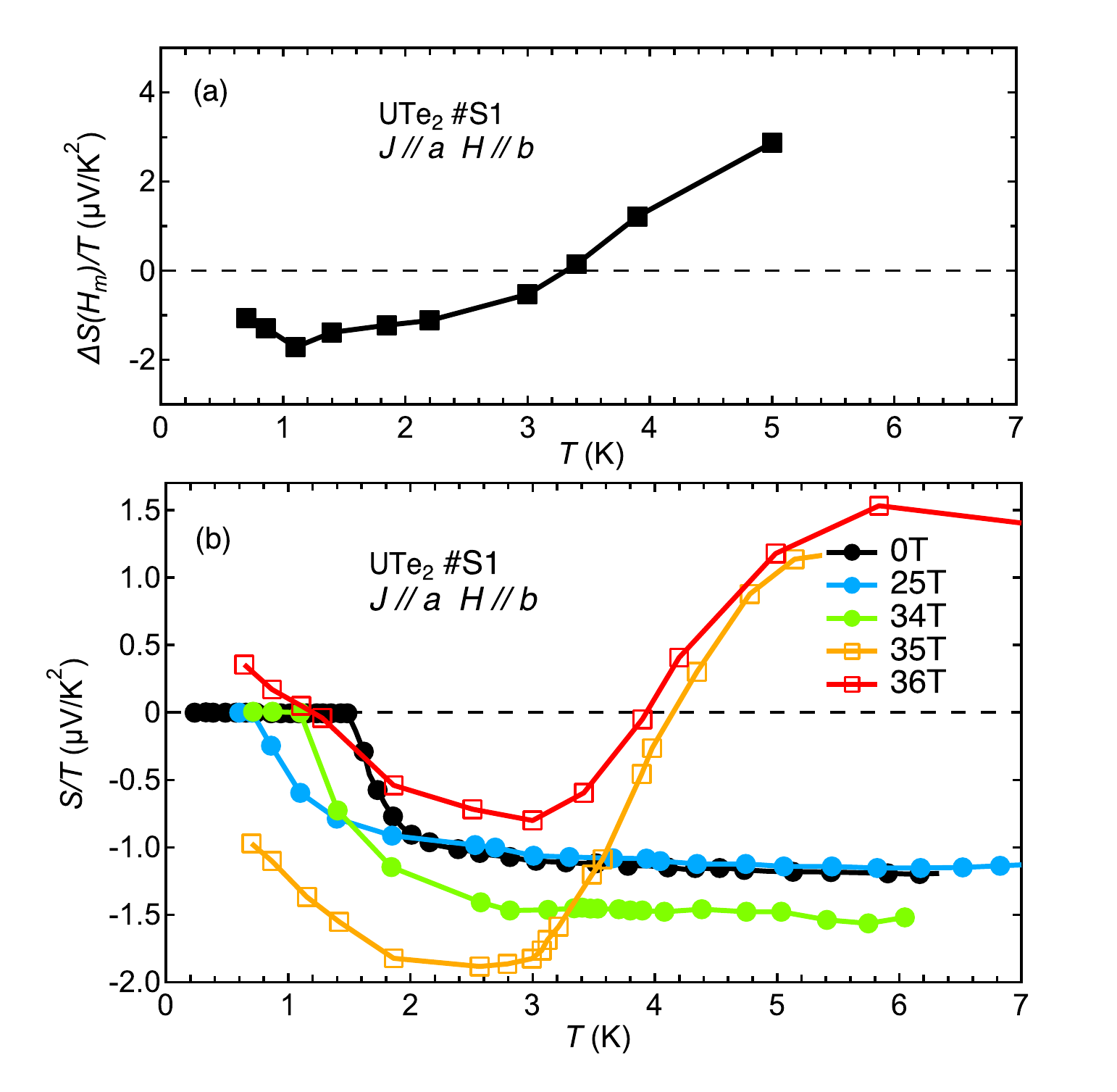}
\caption{(color online) (a) Magnitude of the jump $\Delta S/T=\frac{S(H_\mathrm{m}+ \delta H)-S(H_\mathrm{m}-\delta H)}{T}$ in $S/T$ at $H_\mathrm{m}$. (b) Temperature dependence of $S/T$ at different magnetic fields (full symbols $H<H_m$, open symbols $H>H_m$). $S/T$ changes drastically with temperature near $H_\mathrm{m}$. }
\label{Fig2}
\end{figure}

Figure~\ref{Fig1} shows the magnetic field dependence of $S$ from 0 to 36~T (a) and from 32 to 36~T (b-c) at various temperatures. At 0.7~K, $S$ is equal to zero up to the first order transition at $H_\mathrm{m}=34.6$~T, where $S$ shows a clear negative jump followed by a rapid increase in agreement with the collapse of SC above $H_\mathrm{m}$. At slightly higher temperature 0.86~K, the sample enters the normal state at about 13~T with a negative $S$, as indicated by the arrows in the inset of Fig.~\ref{Fig1}~(a). A field-induced RSC phase is then observed between 27~T and 34.6~T. The first order character of the transition was also observed in our $S(H)$ measurements with a strong hysteresis, (see Fig.~S1 of the Supplemental Material). Upon warming, the hysteresis closes and the jump vanishes, indicating that the first-order transition terminates at a CEP with $T_\mathrm{CEP} \approx7$~K in agreement previous measurements \cite{Miyake2019, Knafo2019}. Below $T_\mathrm{CEP}$, there is a slight increase of  $|S(H)|$ on approaching $H_\mathrm{m}$, then $S(H)$ changes abruptly at $H_\mathrm{m}$. Interestingly, the negative jump at $H_\mathrm{m}$ disappears at 3.4~K and it becomes positive at higher temperatures, as shown more clearly in Fig.~\ref{Fig2}~(a), where we plot $\Delta S(H_\mathrm{m})/T$ as a function of temperature up to 5~K. The amplitude of the transition increases  up to $\approx 3 \mu$V/K$^2$ at 5~K. For $T=10~K$, above $T_\mathrm{CEP}$ only a large crossover can be detected.

Fig.~\ref{Fig2}~(b) shows the temperature dependence of $S(T)/T$ between 0~K and 7~K for different magnetic fields. For $H<H_\mathrm{m}$ (full symbols), $S(T)/T \approx1\mu$V/K$^2$ is temperature independent in the normal state in this temperature range and for field below 24~T.  $|S(T)/T|$ is slightly larger on approaching $H_\mathrm{m}$ as shown for 34~T. In contrast, $S(T)/T$ displays a very different temperature dependence above $H_\mathrm{m}$ (open symbols). For instance, at 35~T, $S/T( T)<0$ at low temperature (SC is suppressed) and decreases up to 3~K. Above 3~K, it increases drastically and  changes sign at around 4.1~K. Moreover, at 36~T, the interesting feature is that $S/T$ becomes positive at very low temperature  below 1.4~K. This shows that at low temperature the sign of the dominant heat carriers changes through $H_\mathrm{m}$ from electrons to holes.



\begin{figure}
\includegraphics[width=0.5\textwidth]{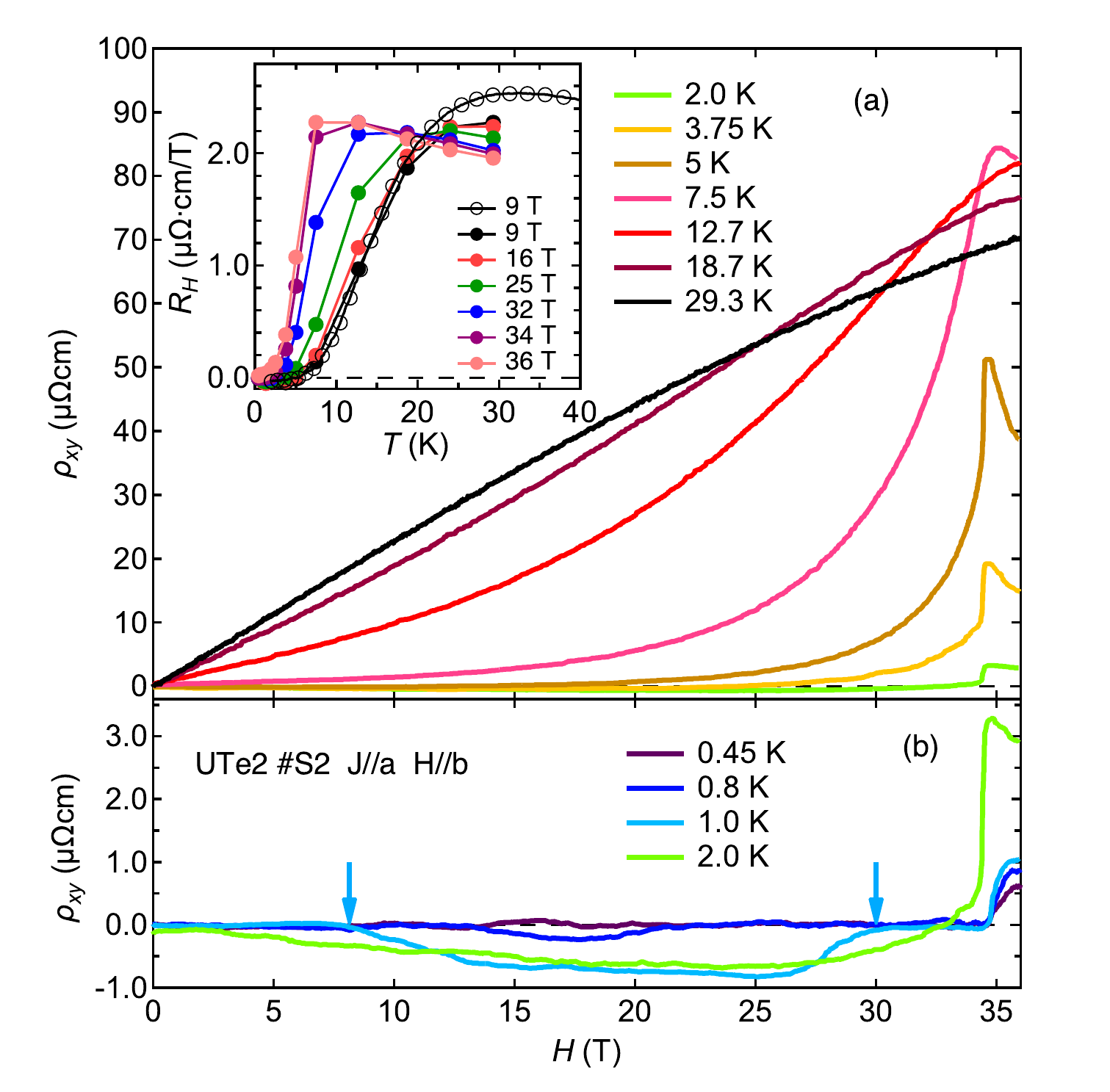}
\caption{(color online) (a) The Hall resistivity of UTe$_2$ with $H \parallel b$ up to 36~T at different temperatures. Panel (b) shows a zoom on low temperatures. The arrows indicate the normal state and RSC at 1~K. The inset of (a) shows the $T$ dependence of the Hall coefficient at different field. Previous data are also represented (open circles) \cite{Niu2019}.}
\label{Fig03}
\end{figure}

To extract more information about the  field dependence of the carriers across $H_\mathrm{m}$, we measured the Hall effect up to 36~T on sample S2 in static field and up to 68~T on sample S3  in pulsed field. Figure~\ref{Fig03}(a) and (b) show the $H$ dependence of the Hall resistivity $\rho_{xy}$ measured on S2 at different temperatures. At 0.45~K, the sample is superconducting up to the MMT. Above $H_\mathrm{m}$,
$\rho_{xy}$ is positive in the normal state.  At 1.0~K, RSC is detected in $\rho_{xy}$ as indicated by the arrows. However, a negative $\rho_{xy}$ shows up in the normal state below the RSC. As the temperature increases, the transition in $\rho_{xy}$ at $H_\mathrm{m}$ becomes huge with a maximum value at 7.5~K near the CEP. At the same time, the initial slope of $\rho_{xy}(H)$ at low field also increases rapidly from negative to positive. The inset of Fig.~\ref{Fig03}~(a) illustrates the temperature dependence of  the Hall coefficient $R_\mathrm{H}=\rho_{xy}/H$ at different fields. At 9~T, where $\rho_{xy}(H)$ is still linear, $R_\mathrm{H}(T)$ changes rapidly from negative to positive and shows a maximum at $T_\mathrm{m}\approx30$~K close to $T_{\chi_\mathrm{max}}$ \cite{Niu2019}. This drastic increase of the Hall coefficient at low temperature, which has been observed in many heavy fermion systems, like UPt$_3$ \cite{Schoenes1986} and UAl$_2$ \cite{HadzicLeroux1986}, is related to the change of the scattering process from incoherent skew scattering at high temperature to a coherent scattering regime at low temperature \cite{Yang2013}. As $H$ increases,  $R_\mathrm{H}(T)$ becomes steeper and $T_\mathrm{m}$ shifts to lower temperature until the MMT transition at $H_\mathrm{m}$, where $T_\mathrm{m}$ ends at about 7~K close to the CEP  (see also Fig.~\ref{Fig05}). Above $H_\mathrm{m}$, $R_\mathrm{H}$ decreases with field and the temperature position of the maximum of $R_\mathrm{H}$ (labelled $T_\mathrm{cr}$) is a signature of the PM-PPM crossover \cite{Gourgout2016,PalacioMorales2015}. It increases to higher temperature when increasing magnetic field (see Fig. S4 of the Supplemental Material). 


In presence of magnetic fluctuations $R_\mathrm{H}$ can be described by the sum of an ordinary part $R_0$ and an anomalous part $R_\mathrm{S}$. $R_0$ is simply related to the density and the mobility of the carriers while $R_\mathrm{S}$ is the result of different scattering processes. In heavy fermion systems, incoherent skew scattering of conduction electrons by independent local $f$ moments predominates at high temperature above the coherence temperature \cite{Fert}. $R_\mathrm{S}$ is proportional to the susceptibility $\chi$ and electrical resistivity $\rho_{xx}$, i.e. $R_\mathrm{S}\propto\rho_{xx}\chi$. This has been verified in many materials in the high temperature incoherent regime. When the coherence settles in at low temperature, a different scattering mechanism, $R_\mathrm{S}\propto\rho_{xx}^2\chi$ has been observed in many uranium heavy fermion compounds \cite{HadzicLeroux1986} and this is theoretically explained by coherent skew scattering \cite{Yamada1993}. 

\begin{figure}
\includegraphics[width=0.5\textwidth]{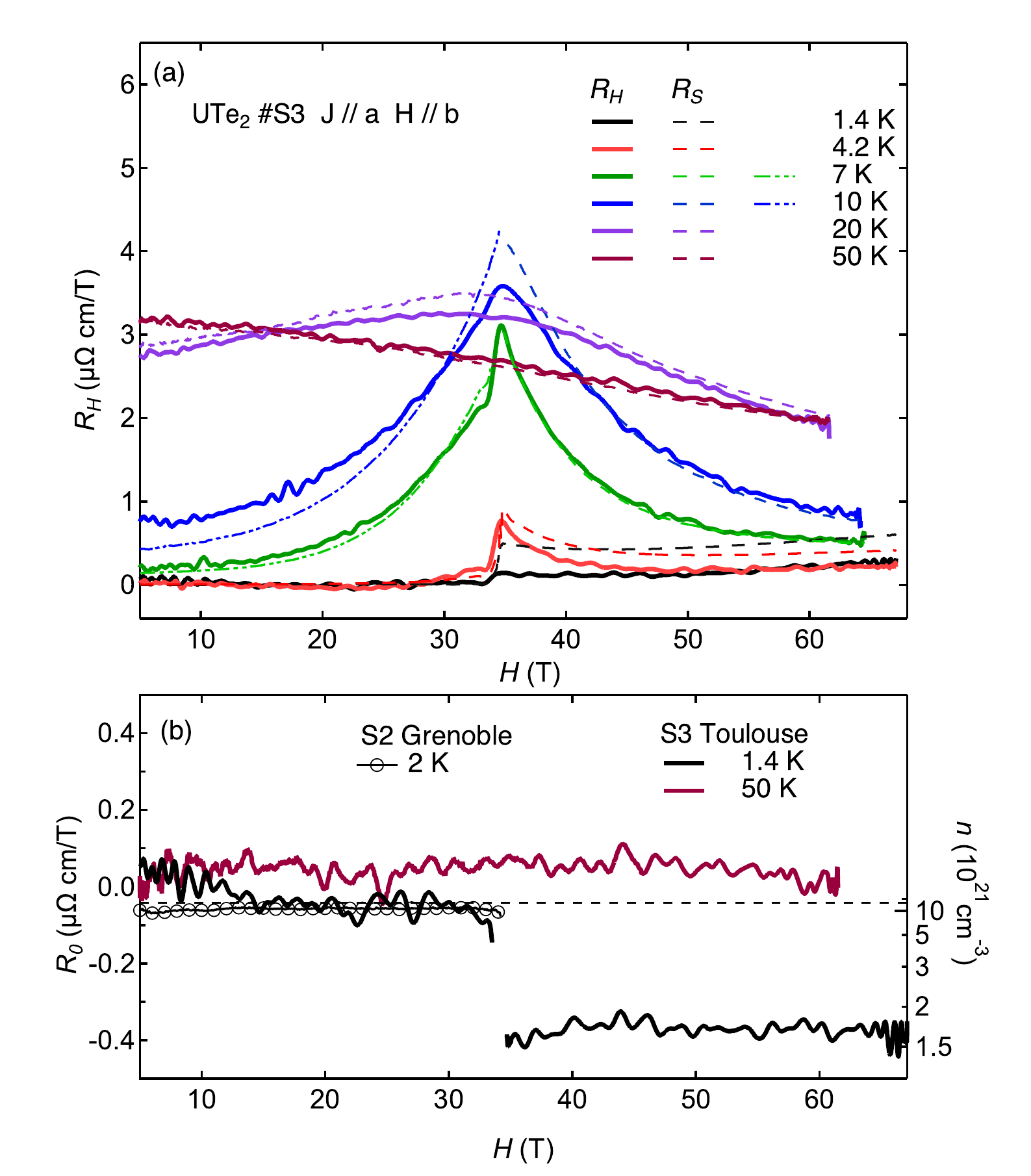}
\caption{(color online) (a) The Hall coefficient of UTe$_2$ S3 with $H \parallel b$ up to 68~T (pulsed field). The dashed lines are estimated from the coherent ($T<20$~K) and incoherent ($T \geqslant 20$~K) skew scattering at different temperatures respectively.  For $T<20$~K, the long and short dashes correspond to the anomalous Hall signal obtained above and below $H_\mathrm{m}$ respectively. (b) $R_0$ is obtained after subtracting the anomalous part from the Hall signal (additional temperature are shown on Fig.~S9). The right scale indicates the carrier density and the dashed line represents the value obtained previously \cite{Niu2019}.}
\label{Fig04}
\end{figure}

In order to get information on the change of carriers in UTe$_2$ (for details see Supplemental Material), we have used susceptibility data of UTe$_2$ from Ref. \onlinecite{Miyake2019} and plotted $R_{xy}/H$ against $R_{xx}M/H$ or $R_{xx}^2M/H$ at different temperatures (see Fig.~S6, S7 and S8).  $R_{xy}/H$ is linear against $R^2_{xx}M/H$ up to $\sim$10~K. However, the curves below and above $H_\mathrm{m}$ fall onto two different lines with different slopes and/or intercepts, indicating that both $R_0$ and $R_\mathrm{S}$ have discontinuous changes at $H_\mathrm{m}$. In contrast on, at 50~K, which is above $T_\mathrm{m}$, the coherence temperature, $R_{xy}/H$ is on a straight line with $R_{xx}M/H$ in almost the whole field range, consistent with the incoherent skew scattering predictions.
The analysis shows that the Hall effect below 10~K is dominated by coherent and above 10~K by incoherent skex scattering. This allows us to estimate the contribution of $R_0$ to the total Hall effect. The solid lines in Fig.~\ref{Fig04}~(a) show $R_{xy}/H$, while the dashed lines correspond to the anomalous Hall contribution obtained from the fitting by considering the change of slope of the anomalous Hall effect below (short dash) and above (long dash) $H_\mathrm{m}$. The difference of these two datasets gives an estimation of $R_0$ as plotted in Fig.~\ref{Fig04}~(b). At 1.4~K, below $H_\mathrm{m}$, $R_\mathrm{0}$ reflects the fact  that the anomalous Hall effect vanishes at low temperature (see also Fig. S5). $R_\mathrm{0}$ is negative, very small, and independent of magnetic field up to $H_\mathrm{m}$. The value of the extracted carrier density (right scale) is in good agreement with the value obtained previously (dashed line) \cite{Niu2019}, $n=1.6 \times 10^{22}$ cm$^{-3}$. Above $H_\mathrm{m}$, $|R_0|$ is much larger and still field independent. Most likely, such a behaviour is the signature of a change of the carrier density (accompanied or not with a change of the mobility)  at the MMT. In contrast, at 50~K entering in the incoherent regime, the Hall coefficient can be very well reproduced by anomalous Hall terms.

\begin{figure}
\includegraphics[width=0.5\textwidth]{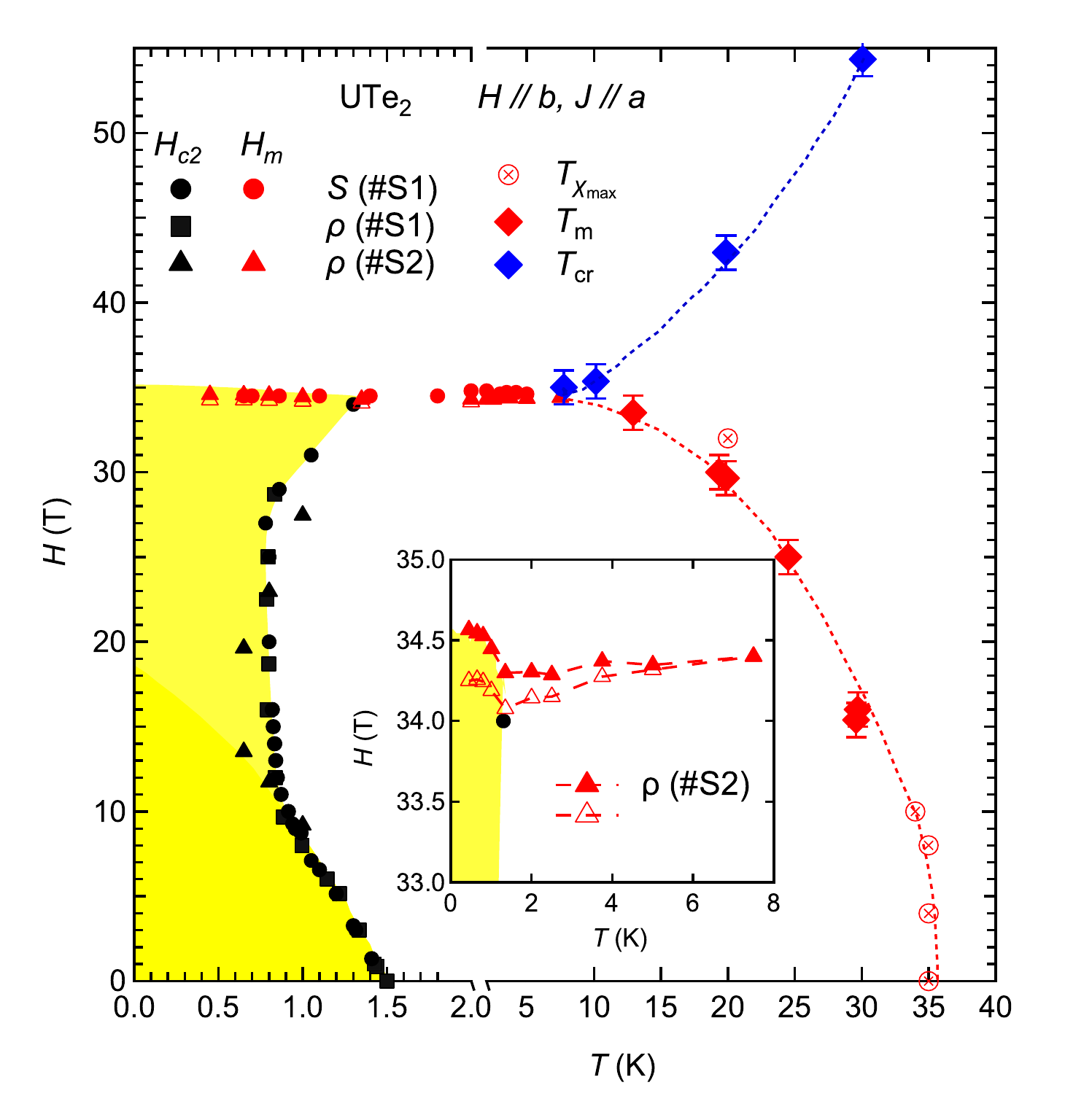}
\caption{(color online) $H-T$ phase diagram of UTe$_2$ for $H \parallel b$. $H_\mathrm{c2}$ from different samples by black symbols and the metamagnetic field $H_\mathrm{m}$ was labelled by red circles ($S$) and red triangles ($\rho$). Red crosses, red diamonds and blue diamonds indicate $T_{\chi_\mathrm{max}}$, $T_\mathrm{m}$ and $T_\mathrm{cr}$ respectively. The inset shows a zoom very close to $H_\mathrm{m}$. Full symbols are from increasing field sweeps, open symbols from downward field sweeps. SC survives slightly above the MMT.}
\label{Fig05}
\end{figure}

$H_\mathrm{c2}(T)$ and $H_\mathrm{m}$ detected in $S$ and $\rho$ are summarised in the $H$-$T$ phase diagram in Fig. \ref{Fig05}. $H_\mathrm{c2}(T)$ is defined by $S=0$ or $\rho=0$. Above 10~T, $S$ and $\rho$ show an almost vertical $H_\mathrm{c2}(T)$ between 10~T and 28~T and $H_\mathrm{c2}$ is strongly enhanced, reaching 1.4~K at $H_\mathrm{m}$, consistent with previous resistivity experiments \cite{Ran2018,Knebel2019}.  The temperatures of the maximum of the Hall effect, $T_\mathrm{m}$ (similar to   $T_{\chi_\mathrm{max}}$) and $T_\mathrm{cr}$ are also represented. This is similar to the energy scales observed near the tri-critical point in the ferromagnetic superconductor URhGe with the same crossover line separating PM and PPM states \cite{Gourgout2016}. Similar TEP experiments in the nearly ferromagnetic (FM) case of  UCoAl \cite{Palacio-Morales2013} have identified the CEP of the first order transition at the MMT from PM to FM states. This material is an itinerant Ising system, where the MMT occurs with $H$ along the Ising axis. In contrast, in UTe$_2$ as well as in URhGe, the MMT occurs for field along the hard magnetization axis. In both systems the MMT is strongly connected to the field enhancement of SC. The inset magnifies the phase diagram near the MMT transition. We observe that $H_{\rm m}$ ($H_{\rm c2}$) has an upturn below 1~ K. This new feature indicates that SC persists above the extrapolation of $H_\mathrm{m}$ to $T=0$~K, although in a very narrow field range.

In conclusion, we have studied the thermoelectric power and the Hall effect of PM superconductor $\mathrm{UTe}_2$ up to 36~T and 68~T respectively, with magnetic field along hard magnetization $b$-axis. RSC was observed in both $S$ and $R_\mathrm{H}$ at low temperature on approaching the MMT. $R_\mathrm{H}$ is  very well described by incoherent skew scattering above the coherence temperature $T_M$ which corresponds roughly to $T_{\chi_\mathrm{max}}$ and coherent skew scattering at lower temperature. The correspondence of these two energy scale highlights the dual character, localized-itinerant, of the f electrons in uranium compounds. Below this Kondo coherence temperature, DFT calculations reveal the emergence of a band structure with a small peak in the density of states at the Fermi level \cite{Miao2020}. The field dependence of $R_\mathrm{H}$ suggests the suppression of  well defined coherent scattering by strong magnetic fluctuations near the MMT. Above $T_\mathrm{SC}$, for $H<H_\mathrm{m}$, the anomalous part of the Hall signal vanishes, opposite to the case $H>H_\mathrm{m}$ where the anomalous part is still present. The strong change of $R_0$ and $S$ evidence a Fermi-surface reconstruction at $H_\mathrm{m}$. As this Fermi surface change leads to the vanishing of SC above $H_\mathrm{m}$, a drastic change in the nature of the SC pairing may happen, contrasting with rather symmetrical variation of  $\gamma(H)$ on crossing $H_\mathrm{m}$.  A still open question is the anomalous SC phase which persists at 30$^{\circ}$ from the $b$ axis in the PPM states above $H_\mathrm{m}$. The originality of UTe$_2$ consists in its proximity to a Kondo-lattice metal-insulator instability which deserves to be studied further under both magnetic field and pressure.

\begin{acknowledgments}
The authors thank K. Izawa, S. Hosoi, H. Harima, J. Ishizuka, Y. Yanase, for stimulating discussions. We acknowledge A. Miyake for high magnetic field magnetization data. This work has been supported by the Universit{\'e} Grenoble Alpes and KAKENHI (JP15H0082, JP15H0084, JP15K21732, JP19H00646, JP16H04006, JP15H05745). We acknowledge support of the LNCMI-CNRS, member the European Magnetic Field Laboratory (EMFL).
\end{acknowledgments}

\bibliographystyle{apsrev4-1}	

\begin{thebibliography}{32}%
\makeatletter
\providecommand \@ifxundefined [1]{%
 \@ifx{#1\undefined}
}%
\providecommand \@ifnum [1]{%
 \ifnum #1\expandafter \@firstoftwo
 \else \expandafter \@secondoftwo
 \fi
}%
\providecommand \@ifx [1]{%
 \ifx #1\expandafter \@firstoftwo
 \else \expandafter \@secondoftwo
 \fi
}%
\providecommand \natexlab [1]{#1}%
\providecommand \enquote  [1]{``#1''}%
\providecommand \bibnamefont  [1]{#1}%
\providecommand \bibfnamefont [1]{#1}%
\providecommand \citenamefont [1]{#1}%
\providecommand \href@noop [0]{\@secondoftwo}%
\providecommand \href [0]{\begingroup \@sanitize@url \@href}%
\providecommand \@href[1]{\@@startlink{#1}\@@href}%
\providecommand \@@href[1]{\endgroup#1\@@endlink}%
\providecommand \@sanitize@url [0]{\catcode `\\12\catcode `\$12\catcode
  `\&12\catcode `\#12\catcode `\^12\catcode `\_12\catcode `\%12\relax}%
\providecommand \@@startlink[1]{}%
\providecommand \@@endlink[0]{}%
\providecommand \url  [0]{\begingroup\@sanitize@url \@url }%
\providecommand \@url [1]{\endgroup\@href {#1}{\urlprefix }}%
\providecommand \urlprefix  [0]{URL }%
\providecommand \Eprint [0]{\href }%
\providecommand \doibase [0]{http://dx.doi.org/}%
\providecommand \selectlanguage [0]{\@gobble}%
\providecommand \bibinfo  [0]{\@secondoftwo}%
\providecommand \bibfield  [0]{\@secondoftwo}%
\providecommand \translation [1]{[#1]}%
\providecommand \BibitemOpen [0]{}%
\providecommand \bibitemStop [0]{}%
\providecommand \bibitemNoStop [0]{.\EOS\space}%
\providecommand \EOS [0]{\spacefactor3000\relax}%
\providecommand \BibitemShut  [1]{\csname bibitem#1\endcsname}%
\let\auto@bib@innerbib\@empty
\bibitem [{\citenamefont {Ran}\ \emph {et~al.}(2019{\natexlab{a}})\citenamefont
  {Ran}, \citenamefont {Eckberg}, \citenamefont {Ding}, \citenamefont
  {Furukawa}, \citenamefont {Metz}, \citenamefont {Saha}, \citenamefont {Liu},
  \citenamefont {Zic}, \citenamefont {Kim}, \citenamefont {Paglione},\ and\
  \citenamefont {Butch}}]{Ran2018}%
  \BibitemOpen
  \bibfield  {author} {\bibinfo {author} {\bibfnamefont {S.}~\bibnamefont
  {Ran}}, \bibinfo {author} {\bibfnamefont {C.}~\bibnamefont {Eckberg}},
  \bibinfo {author} {\bibfnamefont {Q.-P.}\ \bibnamefont {Ding}}, \bibinfo
  {author} {\bibfnamefont {Y.}~\bibnamefont {Furukawa}}, \bibinfo {author}
  {\bibfnamefont {T.}~\bibnamefont {Metz}}, \bibinfo {author} {\bibfnamefont
  {S.~R.}\ \bibnamefont {Saha}}, \bibinfo {author} {\bibfnamefont {I.-L.}\
  \bibnamefont {Liu}}, \bibinfo {author} {\bibfnamefont {M.}~\bibnamefont
  {Zic}}, \bibinfo {author} {\bibfnamefont {H.}~\bibnamefont {Kim}}, \bibinfo
  {author} {\bibfnamefont {J.}~\bibnamefont {Paglione}}, \ and\ \bibinfo
  {author} {\bibfnamefont {N.~P.}\ \bibnamefont {Butch}},\ }\href {\doibase
  10.1126/science.aav8645} {\bibfield  {journal} {\bibinfo  {journal}
  {Science}\ }\textbf {\bibinfo {volume} {365}},\ \bibinfo {pages} {684}
  (\bibinfo {year} {2019}{\natexlab{a}})}\BibitemShut {NoStop}%
\bibitem [{\citenamefont {Aoki}\ \emph {et~al.}(2019)\citenamefont {Aoki},
  \citenamefont {Nakamura}, \citenamefont {Honda}, \citenamefont {Li},
  \citenamefont {Homma}, \citenamefont {Shimizu}, \citenamefont {Sato},
  \citenamefont {Knebel}, \citenamefont {Brison}, \citenamefont {Pourret},
  \citenamefont {Braithwaite}, \citenamefont {Lapertot}, \citenamefont {Niu},
  \citenamefont {Vali{\v{s}}ka}, \citenamefont {Harima},\ and\ \citenamefont
  {Flouquet}}]{Aoki2019}%
  \BibitemOpen
  \bibfield  {author} {\bibinfo {author} {\bibfnamefont {D.}~\bibnamefont
  {Aoki}}, \bibinfo {author} {\bibfnamefont {A.}~\bibnamefont {Nakamura}},
  \bibinfo {author} {\bibfnamefont {F.}~\bibnamefont {Honda}}, \bibinfo
  {author} {\bibfnamefont {D.}~\bibnamefont {Li}}, \bibinfo {author}
  {\bibfnamefont {Y.}~\bibnamefont {Homma}}, \bibinfo {author} {\bibfnamefont
  {Y.}~\bibnamefont {Shimizu}}, \bibinfo {author} {\bibfnamefont {Y.~J.}\
  \bibnamefont {Sato}}, \bibinfo {author} {\bibfnamefont {G.}~\bibnamefont
  {Knebel}}, \bibinfo {author} {\bibfnamefont {J.-P.}\ \bibnamefont {Brison}},
  \bibinfo {author} {\bibfnamefont {A.}~\bibnamefont {Pourret}}, \bibinfo
  {author} {\bibfnamefont {D.}~\bibnamefont {Braithwaite}}, \bibinfo {author}
  {\bibfnamefont {G.}~\bibnamefont {Lapertot}}, \bibinfo {author}
  {\bibfnamefont {Q.}~\bibnamefont {Niu}}, \bibinfo {author} {\bibfnamefont
  {M.}~\bibnamefont {Vali{\v{s}}ka}}, \bibinfo {author} {\bibfnamefont
  {H.}~\bibnamefont {Harima}}, \ and\ \bibinfo {author} {\bibfnamefont
  {J.}~\bibnamefont {Flouquet}},\ }\href {\doibase 10.7566/JPSJ.88.043702}
  {\bibfield  {journal} {\bibinfo  {journal} {J. Phys. Soc. Jpn.}\ }\textbf
  {\bibinfo {volume} {88}},\ \bibinfo {pages} {043702} (\bibinfo {year}
  {2019})}\BibitemShut {NoStop}%
\bibitem [{\citenamefont {Knebel}\ \emph {et~al.}(2019)\citenamefont {Knebel},
  \citenamefont {Knafo}, \citenamefont {Pourret}, \citenamefont {Niu},
  \citenamefont {Vali{\v{s}}ka}, \citenamefont {Braithwaite}, \citenamefont
  {Lapertot}, \citenamefont {Nardone}, \citenamefont {Zitouni}, \citenamefont
  {Mishra}, \citenamefont {Sheikin}, \citenamefont {Seyfarth}, \citenamefont
  {Brison}, \citenamefont {Aoki},\ and\ \citenamefont {Flouquet}}]{Knebel2019}%
  \BibitemOpen
  \bibfield  {author} {\bibinfo {author} {\bibfnamefont {G.}~\bibnamefont
  {Knebel}}, \bibinfo {author} {\bibfnamefont {W.}~\bibnamefont {Knafo}},
  \bibinfo {author} {\bibfnamefont {A.}~\bibnamefont {Pourret}}, \bibinfo
  {author} {\bibfnamefont {Q.}~\bibnamefont {Niu}}, \bibinfo {author}
  {\bibfnamefont {M.}~\bibnamefont {Vali{\v{s}}ka}}, \bibinfo {author}
  {\bibfnamefont {D.}~\bibnamefont {Braithwaite}}, \bibinfo {author}
  {\bibfnamefont {G.}~\bibnamefont {Lapertot}}, \bibinfo {author}
  {\bibfnamefont {M.}~\bibnamefont {Nardone}}, \bibinfo {author} {\bibfnamefont
  {A.}~\bibnamefont {Zitouni}}, \bibinfo {author} {\bibfnamefont
  {S.}~\bibnamefont {Mishra}}, \bibinfo {author} {\bibfnamefont
  {I.}~\bibnamefont {Sheikin}}, \bibinfo {author} {\bibfnamefont
  {G.}~\bibnamefont {Seyfarth}}, \bibinfo {author} {\bibfnamefont {J.-P.}\
  \bibnamefont {Brison}}, \bibinfo {author} {\bibfnamefont {D.}~\bibnamefont
  {Aoki}}, \ and\ \bibinfo {author} {\bibfnamefont {J.}~\bibnamefont
  {Flouquet}},\ }\href {\doibase 10.7566/JPSJ.88.063707} {\bibfield  {journal}
  {\bibinfo  {journal} {J. Phys. Soc. Jpn.}\ }\textbf {\bibinfo {volume}
  {88}},\ \bibinfo {pages} {063707} (\bibinfo {year} {2019})}\BibitemShut
  {NoStop}%
\bibitem [{\citenamefont {Niu}\ \emph {et~al.}(2020)\citenamefont {Niu},
  \citenamefont {Knebel}, \citenamefont {Braithwaite}, \citenamefont {Aoki},
  \citenamefont {Lapertot}, \citenamefont {Seyfarth}, \citenamefont {Brison},
  \citenamefont {Flouquet},\ and\ \citenamefont {Pourret}}]{Niu2019}%
  \BibitemOpen
  \bibfield  {author} {\bibinfo {author} {\bibfnamefont {Q.}~\bibnamefont
  {Niu}}, \bibinfo {author} {\bibfnamefont {G.}~\bibnamefont {Knebel}},
  \bibinfo {author} {\bibfnamefont {D.}~\bibnamefont {Braithwaite}}, \bibinfo
  {author} {\bibfnamefont {D.}~\bibnamefont {Aoki}}, \bibinfo {author}
  {\bibfnamefont {G.}~\bibnamefont {Lapertot}}, \bibinfo {author}
  {\bibfnamefont {G.}~\bibnamefont {Seyfarth}}, \bibinfo {author}
  {\bibfnamefont {J.-P.}\ \bibnamefont {Brison}}, \bibinfo {author}
  {\bibfnamefont {J.}~\bibnamefont {Flouquet}}, \ and\ \bibinfo {author}
  {\bibfnamefont {A.}~\bibnamefont {Pourret}},\ }\href {\doibase
  10.1103/PhysRevLett.124.086601} {\bibfield  {journal} {\bibinfo  {journal}
  {Phys. Rev. Lett.}\ }\textbf {\bibinfo {volume} {124}},\ \bibinfo {pages}
  {086601} (\bibinfo {year} {2020})}\BibitemShut {NoStop}%
\bibitem [{\citenamefont {Ishizuka}\ \emph {et~al.}(2019)\citenamefont
  {Ishizuka}, \citenamefont {Sumita}, \citenamefont {Daido},\ and\
  \citenamefont {Yanase}}]{Ishizuka2019}%
  \BibitemOpen
  \bibfield  {author} {\bibinfo {author} {\bibfnamefont {J.}~\bibnamefont
  {Ishizuka}}, \bibinfo {author} {\bibfnamefont {S.}~\bibnamefont {Sumita}},
  \bibinfo {author} {\bibfnamefont {A.}~\bibnamefont {Daido}}, \ and\ \bibinfo
  {author} {\bibfnamefont {Y.}~\bibnamefont {Yanase}},\ }\href {\doibase
  10.1103/PhysRevLett.123.217001} {\bibfield  {journal} {\bibinfo  {journal}
  {Phys. Rev. Lett.}\ }\textbf {\bibinfo {volume} {123}},\ \bibinfo {pages}
  {217001} (\bibinfo {year} {2019})}\BibitemShut {NoStop}%
\bibitem [{\citenamefont {Fujimori}\ \emph {et~al.}(2019)\citenamefont
  {Fujimori}, \citenamefont {Kawasaki}, \citenamefont {Takeda}, \citenamefont
  {Yamagami}, \citenamefont {Nakamura}, \citenamefont {Homma},\ and\
  \citenamefont {Aoki}}]{Fujimori2019}%
  \BibitemOpen
  \bibfield  {author} {\bibinfo {author} {\bibfnamefont {S.-i.}\ \bibnamefont
  {Fujimori}}, \bibinfo {author} {\bibfnamefont {I.}~\bibnamefont {Kawasaki}},
  \bibinfo {author} {\bibfnamefont {Y.}~\bibnamefont {Takeda}}, \bibinfo
  {author} {\bibfnamefont {H.}~\bibnamefont {Yamagami}}, \bibinfo {author}
  {\bibfnamefont {A.}~\bibnamefont {Nakamura}}, \bibinfo {author}
  {\bibfnamefont {Y.}~\bibnamefont {Homma}}, \ and\ \bibinfo {author}
  {\bibfnamefont {D.}~\bibnamefont {Aoki}},\ }\href {\doibase
  10.7566/JPSJ.88.103701} {\bibfield  {journal} {\bibinfo  {journal} {J. Phys.
  Soc. Jpn.}\ }\textbf {\bibinfo {volume} {88}},\ \bibinfo {pages} {103701}
  (\bibinfo {year} {2019})}\BibitemShut {NoStop}%
\bibitem [{\citenamefont {Xu}\ \emph {et~al.}(2019)\citenamefont {Xu},
  \citenamefont {Sheng},\ and\ \citenamefont {Yang}}]{Xu2019}%
  \BibitemOpen
  \bibfield  {author} {\bibinfo {author} {\bibfnamefont {Y.}~\bibnamefont
  {Xu}}, \bibinfo {author} {\bibfnamefont {Y.}~\bibnamefont {Sheng}}, \ and\
  \bibinfo {author} {\bibfnamefont {Y.~F.}\ \bibnamefont {Yang}},\ }\href
  {\doibase 10.1103/PhysRevLett.123.217002} {\bibfield  {journal} {\bibinfo
  {journal} {Phys. Rev. Lett.}\ }\textbf {\bibinfo {volume} {123}},\ \bibinfo
  {pages} {217002} (\bibinfo {year} {2019})}\BibitemShut {NoStop}%
\bibitem [{\citenamefont {Miao}\ \emph
  {et~al.}(2020{\natexlab{a}})\citenamefont {Miao}, \citenamefont {Liu},
  \citenamefont {Xu}, \citenamefont {Kotta}, \citenamefont {Kang},
  \citenamefont {Ran}, \citenamefont {Paglione}, \citenamefont {Kotliar},
  \citenamefont {Butch}, \citenamefont {Denlinger},\ and\ \citenamefont
  {Wray}}]{Miao2019}%
  \BibitemOpen
  \bibfield  {author} {\bibinfo {author} {\bibfnamefont {L.}~\bibnamefont
  {Miao}}, \bibinfo {author} {\bibfnamefont {S.}~\bibnamefont {Liu}}, \bibinfo
  {author} {\bibfnamefont {Y.}~\bibnamefont {Xu}}, \bibinfo {author}
  {\bibfnamefont {E.~C.}\ \bibnamefont {Kotta}}, \bibinfo {author}
  {\bibfnamefont {C.-J.}\ \bibnamefont {Kang}}, \bibinfo {author}
  {\bibfnamefont {S.}~\bibnamefont {Ran}}, \bibinfo {author} {\bibfnamefont
  {J.}~\bibnamefont {Paglione}}, \bibinfo {author} {\bibfnamefont
  {G.}~\bibnamefont {Kotliar}}, \bibinfo {author} {\bibfnamefont {N.~P.}\
  \bibnamefont {Butch}}, \bibinfo {author} {\bibfnamefont {J.~D.}\ \bibnamefont
  {Denlinger}}, \ and\ \bibinfo {author} {\bibfnamefont {L.~A.}\ \bibnamefont
  {Wray}},\ }\href {\doibase 10.1103/PhysRevLett.124.076401} {\bibfield
  {journal} {\bibinfo  {journal} {Phys. Rev. Lett.}\ }\textbf {\bibinfo
  {volume} {124}},\ \bibinfo {pages} {076401} (\bibinfo {year}
  {2020}{\natexlab{a}})}\BibitemShut {NoStop}%
\bibitem [{\citenamefont {Tokunaga}\ \emph {et~al.}(2019)\citenamefont
  {Tokunaga}, \citenamefont {Sakai}, \citenamefont {Kambe}, \citenamefont
  {Hattori}, \citenamefont {Higa}, \citenamefont {Nakamine}, \citenamefont
  {Kitagawa}, \citenamefont {Ishida}, \citenamefont {Nakamura}, \citenamefont
  {Shimizu}, \citenamefont {Homma}, \citenamefont {Li}, \citenamefont {Honda},\
  and\ \citenamefont {Aoki}}]{Tokunaga2019}%
  \BibitemOpen
  \bibfield  {author} {\bibinfo {author} {\bibfnamefont {Y.}~\bibnamefont
  {Tokunaga}}, \bibinfo {author} {\bibfnamefont {H.}~\bibnamefont {Sakai}},
  \bibinfo {author} {\bibfnamefont {S.}~\bibnamefont {Kambe}}, \bibinfo
  {author} {\bibfnamefont {T.}~\bibnamefont {Hattori}}, \bibinfo {author}
  {\bibfnamefont {N.}~\bibnamefont {Higa}}, \bibinfo {author} {\bibfnamefont
  {G.}~\bibnamefont {Nakamine}}, \bibinfo {author} {\bibfnamefont
  {S.}~\bibnamefont {Kitagawa}}, \bibinfo {author} {\bibfnamefont
  {K.}~\bibnamefont {Ishida}}, \bibinfo {author} {\bibfnamefont
  {A.}~\bibnamefont {Nakamura}}, \bibinfo {author} {\bibfnamefont
  {Y.}~\bibnamefont {Shimizu}}, \bibinfo {author} {\bibfnamefont
  {Y.}~\bibnamefont {Homma}}, \bibinfo {author} {\bibfnamefont
  {D.}~\bibnamefont {Li}}, \bibinfo {author} {\bibfnamefont {F.}~\bibnamefont
  {Honda}}, \ and\ \bibinfo {author} {\bibfnamefont {D.}~\bibnamefont {Aoki}},\
  }\href {\doibase 10.7566/JPSJ.88.073701} {\bibfield  {journal} {\bibinfo
  {journal} {J. Phys. Soc. Jpn.}\ }\textbf {\bibinfo {volume} {88}},\ \bibinfo
  {pages} {073701} (\bibinfo {year} {2019})}\BibitemShut {NoStop}%
\bibitem [{\citenamefont {Ran}\ \emph {et~al.}(2019{\natexlab{b}})\citenamefont
  {Ran}, \citenamefont {Liu}, \citenamefont {Eo}, \citenamefont {Campbell},
  \citenamefont {Neves}, \citenamefont {Fuhrman}, \citenamefont {Saha},
  \citenamefont {Eckberg}, \citenamefont {Kim}, \citenamefont {Graf},
  \citenamefont {Balakirev}, \citenamefont {Singleton}, \citenamefont
  {Paglione},\ and\ \citenamefont {Butch}}]{Ran2019}%
  \BibitemOpen
  \bibfield  {author} {\bibinfo {author} {\bibfnamefont {S.}~\bibnamefont
  {Ran}}, \bibinfo {author} {\bibfnamefont {I.-L.}\ \bibnamefont {Liu}},
  \bibinfo {author} {\bibfnamefont {Y.~S.}\ \bibnamefont {Eo}}, \bibinfo
  {author} {\bibfnamefont {D.~J.}\ \bibnamefont {Campbell}}, \bibinfo {author}
  {\bibfnamefont {P.~M.}\ \bibnamefont {Neves}}, \bibinfo {author}
  {\bibfnamefont {W.~T.}\ \bibnamefont {Fuhrman}}, \bibinfo {author}
  {\bibfnamefont {S.~R.}\ \bibnamefont {Saha}}, \bibinfo {author}
  {\bibfnamefont {C.}~\bibnamefont {Eckberg}}, \bibinfo {author} {\bibfnamefont
  {H.}~\bibnamefont {Kim}}, \bibinfo {author} {\bibfnamefont {D.}~\bibnamefont
  {Graf}}, \bibinfo {author} {\bibfnamefont {F.}~\bibnamefont {Balakirev}},
  \bibinfo {author} {\bibfnamefont {J.}~\bibnamefont {Singleton}}, \bibinfo
  {author} {\bibfnamefont {J.}~\bibnamefont {Paglione}}, \ and\ \bibinfo
  {author} {\bibfnamefont {N.~P.}\ \bibnamefont {Butch}},\ }\href {\doibase
  10.1038/s41567-019-0670-x} {\bibfield  {journal} {\bibinfo  {journal} {Nat.
  Phys.}\ }\textbf {\bibinfo {volume} {15}},\ \bibinfo {pages} {1250} (\bibinfo
  {year} {2019}{\natexlab{b}})}\BibitemShut {NoStop}%
\bibitem [{\citenamefont {Nakamine}\ \emph {et~al.}(2019)\citenamefont
  {Nakamine}, \citenamefont {Kitagawa}, \citenamefont {Ishida}, \citenamefont
  {Tokunaga}, \citenamefont {Sakai}, \citenamefont {Kambe}, \citenamefont
  {Nakamura}, \citenamefont {Shimizu}, \citenamefont {Homma}, \citenamefont
  {Li}, \citenamefont {Honda},\ and\ \citenamefont {Aoki}}]{Nakamine2019}%
  \BibitemOpen
  \bibfield  {author} {\bibinfo {author} {\bibfnamefont {G.}~\bibnamefont
  {Nakamine}}, \bibinfo {author} {\bibfnamefont {S.}~\bibnamefont {Kitagawa}},
  \bibinfo {author} {\bibfnamefont {K.}~\bibnamefont {Ishida}}, \bibinfo
  {author} {\bibfnamefont {Y.}~\bibnamefont {Tokunaga}}, \bibinfo {author}
  {\bibfnamefont {H.}~\bibnamefont {Sakai}}, \bibinfo {author} {\bibfnamefont
  {S.}~\bibnamefont {Kambe}}, \bibinfo {author} {\bibfnamefont
  {A.}~\bibnamefont {Nakamura}}, \bibinfo {author} {\bibfnamefont
  {Y.}~\bibnamefont {Shimizu}}, \bibinfo {author} {\bibfnamefont
  {Y.}~\bibnamefont {Homma}}, \bibinfo {author} {\bibfnamefont
  {D.}~\bibnamefont {Li}}, \bibinfo {author} {\bibfnamefont {F.}~\bibnamefont
  {Honda}}, \ and\ \bibinfo {author} {\bibfnamefont {D.}~\bibnamefont {Aoki}},\
  }\href {\doibase 10.7566/JPSJ.88.113703} {\bibfield  {journal} {\bibinfo
  {journal} {J. Phys. Soc. Jpn.}\ }\textbf {\bibinfo {volume} {88}},\ \bibinfo
  {pages} {113703} (\bibinfo {year} {2019})}\BibitemShut {NoStop}%
\bibitem [{\citenamefont {Miyake}\ \emph {et~al.}(2019)\citenamefont {Miyake},
  \citenamefont {Shimizu}, \citenamefont {Sato}, \citenamefont {Li},
  \citenamefont {Nakamura}, \citenamefont {Homma}, \citenamefont {Honda},
  \citenamefont {Flouquet}, \citenamefont {Tokunaga},\ and\ \citenamefont
  {Aoki}}]{Miyake2019}%
  \BibitemOpen
  \bibfield  {author} {\bibinfo {author} {\bibfnamefont {A.}~\bibnamefont
  {Miyake}}, \bibinfo {author} {\bibfnamefont {Y.}~\bibnamefont {Shimizu}},
  \bibinfo {author} {\bibfnamefont {Y.~J.}\ \bibnamefont {Sato}}, \bibinfo
  {author} {\bibfnamefont {D.}~\bibnamefont {Li}}, \bibinfo {author}
  {\bibfnamefont {A.}~\bibnamefont {Nakamura}}, \bibinfo {author}
  {\bibfnamefont {Y.}~\bibnamefont {Homma}}, \bibinfo {author} {\bibfnamefont
  {F.}~\bibnamefont {Honda}}, \bibinfo {author} {\bibfnamefont
  {J.}~\bibnamefont {Flouquet}}, \bibinfo {author} {\bibfnamefont
  {M.}~\bibnamefont {Tokunaga}}, \ and\ \bibinfo {author} {\bibfnamefont
  {D.}~\bibnamefont {Aoki}},\ }\href {\doibase 10.7566/JPSJ.88.063706}
  {\bibfield  {journal} {\bibinfo  {journal} {J. Phys. Soc. Jpn.}\ }\textbf
  {\bibinfo {volume} {88}},\ \bibinfo {pages} {063706} (\bibinfo {year}
  {2019})}\BibitemShut {NoStop}%
\bibitem [{\citenamefont {Imajo}\ \emph {et~al.}(2019)\citenamefont {Imajo},
  \citenamefont {Kohama}, \citenamefont {Miyake}, \citenamefont {Dong},
  \citenamefont {Tokunaga}, \citenamefont {Flouquet}, \citenamefont {Kindo},\
  and\ \citenamefont {Aoki}}]{Imajo2019}%
  \BibitemOpen
  \bibfield  {author} {\bibinfo {author} {\bibfnamefont {S.}~\bibnamefont
  {Imajo}}, \bibinfo {author} {\bibfnamefont {Y.}~\bibnamefont {Kohama}},
  \bibinfo {author} {\bibfnamefont {A.}~\bibnamefont {Miyake}}, \bibinfo
  {author} {\bibfnamefont {C.}~\bibnamefont {Dong}}, \bibinfo {author}
  {\bibfnamefont {M.}~\bibnamefont {Tokunaga}}, \bibinfo {author}
  {\bibfnamefont {J.}~\bibnamefont {Flouquet}}, \bibinfo {author}
  {\bibfnamefont {K.}~\bibnamefont {Kindo}}, \ and\ \bibinfo {author}
  {\bibfnamefont {D.}~\bibnamefont {Aoki}},\ }\href {\doibase
  10.7566/JPSJ.88.083705} {\bibfield  {journal} {\bibinfo  {journal} {J. Phys.
  Soc. Jpn.}\ }\textbf {\bibinfo {volume} {88}},\ \bibinfo {pages} {083705}
  (\bibinfo {year} {2019})}\BibitemShut {NoStop}%
\bibitem [{\citenamefont {Knafo}\ \emph {et~al.}(2019)\citenamefont {Knafo},
  \citenamefont {Vali{\v{s}}ka}, \citenamefont {Braithwaite}, \citenamefont
  {Lapertot}, \citenamefont {Knebel}, \citenamefont {Pourret}, \citenamefont
  {Brison}, \citenamefont {Flouquet},\ and\ \citenamefont {Aoki}}]{Knafo2019}%
  \BibitemOpen
  \bibfield  {author} {\bibinfo {author} {\bibfnamefont {W.}~\bibnamefont
  {Knafo}}, \bibinfo {author} {\bibfnamefont {M.}~\bibnamefont
  {Vali{\v{s}}ka}}, \bibinfo {author} {\bibfnamefont {D.}~\bibnamefont
  {Braithwaite}}, \bibinfo {author} {\bibfnamefont {G.}~\bibnamefont
  {Lapertot}}, \bibinfo {author} {\bibfnamefont {G.}~\bibnamefont {Knebel}},
  \bibinfo {author} {\bibfnamefont {A.}~\bibnamefont {Pourret}}, \bibinfo
  {author} {\bibfnamefont {J.-P.}\ \bibnamefont {Brison}}, \bibinfo {author}
  {\bibfnamefont {J.}~\bibnamefont {Flouquet}}, \ and\ \bibinfo {author}
  {\bibfnamefont {D.}~\bibnamefont {Aoki}},\ }\href {\doibase
  10.7566/JPSJ.88.063705} {\bibfield  {journal} {\bibinfo  {journal} {J. Phys.
  Soc. Jpn.}\ }\textbf {\bibinfo {volume} {88}},\ \bibinfo {pages} {063705}
  (\bibinfo {year} {2019})}\BibitemShut {NoStop}%
\bibitem [{\citenamefont {Aoki}\ \emph {et~al.}(2001)\citenamefont {Aoki},
  \citenamefont {Huxley}, \citenamefont {Ressouche}, \citenamefont
  {Braithwaite}, \citenamefont {Flouquet}, \citenamefont {Brison},
  \citenamefont {Lhotel},\ and\ \citenamefont {Paulsen}}]{Aoki2001}%
  \BibitemOpen
  \bibfield  {author} {\bibinfo {author} {\bibfnamefont {D.}~\bibnamefont
  {Aoki}}, \bibinfo {author} {\bibfnamefont {A.}~\bibnamefont {Huxley}},
  \bibinfo {author} {\bibfnamefont {E.}~\bibnamefont {Ressouche}}, \bibinfo
  {author} {\bibfnamefont {D.}~\bibnamefont {Braithwaite}}, \bibinfo {author}
  {\bibfnamefont {J.}~\bibnamefont {Flouquet}}, \bibinfo {author}
  {\bibfnamefont {J.~P.}\ \bibnamefont {Brison}}, \bibinfo {author}
  {\bibfnamefont {E.}~\bibnamefont {Lhotel}}, \ and\ \bibinfo {author}
  {\bibfnamefont {C.}~\bibnamefont {Paulsen}},\ }\href {\doibase
  10.1038/35098048} {\bibfield  {journal} {\bibinfo  {journal} {Nature}\
  }\textbf {\bibinfo {volume} {413}},\ \bibinfo {pages} {613} (\bibinfo {year}
  {2001})}\BibitemShut {NoStop}%
\bibitem [{\citenamefont {Miyake}\ \emph {et~al.}(2008)\citenamefont {Miyake},
  \citenamefont {Aoki},\ and\ \citenamefont {Flouquet}}]{Miyake2008a}%
  \BibitemOpen
  \bibfield  {author} {\bibinfo {author} {\bibfnamefont {A.}~\bibnamefont
  {Miyake}}, \bibinfo {author} {\bibfnamefont {D.}~\bibnamefont {Aoki}}, \ and\
  \bibinfo {author} {\bibfnamefont {J.}~\bibnamefont {Flouquet}},\ }\href
  {\doibase 10.1143/JPSJ.77.094709} {\bibfield  {journal} {\bibinfo  {journal}
  {J. Phys. Soc. Jpn.}\ }\textbf {\bibinfo {volume} {77}},\ \bibinfo {pages}
  {094709} (\bibinfo {year} {2008})}\BibitemShut {NoStop}%
\bibitem [{\citenamefont {Hardy}\ \emph {et~al.}(2011)\citenamefont {Hardy},
  \citenamefont {Aoki}, \citenamefont {Meingast}, \citenamefont {Schweiss},
  \citenamefont {Burger}, \citenamefont {v.~L{\"{o}}hneysen},\ and\
  \citenamefont {Flouquet}}]{Hardy2011a}%
  \BibitemOpen
  \bibfield  {author} {\bibinfo {author} {\bibfnamefont {F.}~\bibnamefont
  {Hardy}}, \bibinfo {author} {\bibfnamefont {D.}~\bibnamefont {Aoki}},
  \bibinfo {author} {\bibfnamefont {C.}~\bibnamefont {Meingast}}, \bibinfo
  {author} {\bibfnamefont {P.}~\bibnamefont {Schweiss}}, \bibinfo {author}
  {\bibfnamefont {P.}~\bibnamefont {Burger}}, \bibinfo {author} {\bibfnamefont
  {H.}~\bibnamefont {v.~L{\"{o}}hneysen}}, \ and\ \bibinfo {author}
  {\bibfnamefont {J.}~\bibnamefont {Flouquet}},\ }\href {\doibase
  10.1103/PhysRevB.83.195107} {\bibfield  {journal} {\bibinfo  {journal} {Phys.
  Rev. B}\ }\textbf {\bibinfo {volume} {83}},\ \bibinfo {pages} {195107}
  (\bibinfo {year} {2011})}\BibitemShut {NoStop}%
\bibitem [{\citenamefont {Wu}\ \emph {et~al.}(2017)\citenamefont {Wu},
  \citenamefont {Bastien}, \citenamefont {Taupin}, \citenamefont {Paulsen},
  \citenamefont {Howald}, \citenamefont {Aoki},\ and\ \citenamefont
  {Brison}}]{Wu2017}%
  \BibitemOpen
  \bibfield  {author} {\bibinfo {author} {\bibfnamefont {B.}~\bibnamefont
  {Wu}}, \bibinfo {author} {\bibfnamefont {G.}~\bibnamefont {Bastien}},
  \bibinfo {author} {\bibfnamefont {M.}~\bibnamefont {Taupin}}, \bibinfo
  {author} {\bibfnamefont {C.}~\bibnamefont {Paulsen}}, \bibinfo {author}
  {\bibfnamefont {L.}~\bibnamefont {Howald}}, \bibinfo {author} {\bibfnamefont
  {D.}~\bibnamefont {Aoki}}, \ and\ \bibinfo {author} {\bibfnamefont {J.-P.}\
  \bibnamefont {Brison}},\ }\href {\doibase 10.1038/ncomms14480} {\bibfield
  {journal} {\bibinfo  {journal} {Nat. Commun.}\ }\textbf {\bibinfo {volume}
  {8}},\ \bibinfo {pages} {14480} (\bibinfo {year} {2017})}\BibitemShut
  {NoStop}%
\bibitem [{\citenamefont {Gourgout}\ \emph {et~al.}(2016)\citenamefont
  {Gourgout}, \citenamefont {Pourret}, \citenamefont {Knebel}, \citenamefont
  {Aoki}, \citenamefont {Seyfarth},\ and\ \citenamefont
  {Flouquet}}]{Gourgout2016}%
  \BibitemOpen
  \bibfield  {author} {\bibinfo {author} {\bibfnamefont {A.}~\bibnamefont
  {Gourgout}}, \bibinfo {author} {\bibfnamefont {A.}~\bibnamefont {Pourret}},
  \bibinfo {author} {\bibfnamefont {G.}~\bibnamefont {Knebel}}, \bibinfo
  {author} {\bibfnamefont {D.}~\bibnamefont {Aoki}}, \bibinfo {author}
  {\bibfnamefont {G.}~\bibnamefont {Seyfarth}}, \ and\ \bibinfo {author}
  {\bibfnamefont {J.}~\bibnamefont {Flouquet}},\ }\href {\doibase
  10.1103/PhysRevLett.117.046401} {\bibfield  {journal} {\bibinfo  {journal}
  {Phys. Rev. Lett.}\ }\textbf {\bibinfo {volume} {117}},\ \bibinfo {pages}
  {046401} (\bibinfo {year} {2016})}\BibitemShut {NoStop}%
\bibitem [{\citenamefont {Aoki}\ \emph {et~al.}(2014)\citenamefont {Aoki},
  \citenamefont {Knebel},\ and\ \citenamefont {Flouquet}}]{Aoki2014}%
  \BibitemOpen
  \bibfield  {author} {\bibinfo {author} {\bibfnamefont {D.}~\bibnamefont
  {Aoki}}, \bibinfo {author} {\bibfnamefont {G.}~\bibnamefont {Knebel}}, \ and\
  \bibinfo {author} {\bibfnamefont {J.}~\bibnamefont {Flouquet}},\ }\href
  {\doibase 10.7566/JPSJ.83.094719} {\bibfield  {journal} {\bibinfo  {journal}
  {J. Phys. Soc. Jpn.}\ }\textbf {\bibinfo {volume} {83}},\ \bibinfo {pages}
  {094719} (\bibinfo {year} {2014})}\BibitemShut {NoStop}%
\bibitem [{\citenamefont {Yelland}\ \emph {et~al.}(2011)\citenamefont
  {Yelland}, \citenamefont {Barraclough}, \citenamefont {Wang}, \citenamefont
  {Kamenev},\ and\ \citenamefont {Huxley}}]{Yelland2011}%
  \BibitemOpen
  \bibfield  {author} {\bibinfo {author} {\bibfnamefont {E.~a.}\ \bibnamefont
  {Yelland}}, \bibinfo {author} {\bibfnamefont {J.~M.}\ \bibnamefont
  {Barraclough}}, \bibinfo {author} {\bibfnamefont {W.}~\bibnamefont {Wang}},
  \bibinfo {author} {\bibfnamefont {K.~V.}\ \bibnamefont {Kamenev}}, \ and\
  \bibinfo {author} {\bibfnamefont {a.~D.}\ \bibnamefont {Huxley}},\ }\href
  {\doibase 10.1038/nphys2073} {\bibfield  {journal} {\bibinfo  {journal} {Nat.
  Phys.}\ }\textbf {\bibinfo {volume} {7}},\ \bibinfo {pages} {890} (\bibinfo
  {year} {2011})}\BibitemShut {NoStop}%
\bibitem [{\citenamefont {Sherkunov}\ \emph {et~al.}(2018)\citenamefont
  {Sherkunov}, \citenamefont {Chubukov},\ and\ \citenamefont
  {Betouras}}]{Sherkunov2018}%
  \BibitemOpen
  \bibfield  {author} {\bibinfo {author} {\bibfnamefont {Y.}~\bibnamefont
  {Sherkunov}}, \bibinfo {author} {\bibfnamefont {A.~V.}\ \bibnamefont
  {Chubukov}}, \ and\ \bibinfo {author} {\bibfnamefont {J.~J.}\ \bibnamefont
  {Betouras}},\ }\href {\doibase 10.1103/PhysRevLett.121.097001} {\bibfield
  {journal} {\bibinfo  {journal} {Phys. Rev. Lett.}\ }\textbf {\bibinfo
  {volume} {121}},\ \bibinfo {pages} {097001} (\bibinfo {year}
  {2018})}\BibitemShut {NoStop}%
\bibitem [{\citenamefont {Sundar}\ \emph {et~al.}(2019)\citenamefont {Sundar},
  \citenamefont {Gheidi}, \citenamefont {Akintola}, \citenamefont
  {C{\^{o}}t{\'{e}}}, \citenamefont {Dunsiger}, \citenamefont {Ran},
  \citenamefont {Butch}, \citenamefont {Saha}, \citenamefont {Paglione},\ and\
  \citenamefont {Sonier}}]{Sundar2019}%
  \BibitemOpen
  \bibfield  {author} {\bibinfo {author} {\bibfnamefont {S.}~\bibnamefont
  {Sundar}}, \bibinfo {author} {\bibfnamefont {S.}~\bibnamefont {Gheidi}},
  \bibinfo {author} {\bibfnamefont {K.}~\bibnamefont {Akintola}}, \bibinfo
  {author} {\bibfnamefont {A.~M.}\ \bibnamefont {C{\^{o}}t{\'{e}}}}, \bibinfo
  {author} {\bibfnamefont {S.~R.}\ \bibnamefont {Dunsiger}}, \bibinfo {author}
  {\bibfnamefont {S.}~\bibnamefont {Ran}}, \bibinfo {author} {\bibfnamefont
  {N.~P.}\ \bibnamefont {Butch}}, \bibinfo {author} {\bibfnamefont {S.~R.}\
  \bibnamefont {Saha}}, \bibinfo {author} {\bibfnamefont {J.}~\bibnamefont
  {Paglione}}, \ and\ \bibinfo {author} {\bibfnamefont {J.~E.}\ \bibnamefont
  {Sonier}},\ }\href {\doibase 10.1103/PhysRevB.100.140502} {\bibfield
  {journal} {\bibinfo  {journal} {Phys. Rev. B}\ }\textbf {\bibinfo {volume}
  {100}},\ \bibinfo {pages} {140502} (\bibinfo {year} {2019})}\BibitemShut
  {NoStop}%
\bibitem [{\citenamefont {Paulsen}\ \emph {et~al.}(2020)\citenamefont
  {Paulsen}, \citenamefont {Knebel}, \citenamefont {Lapertot}, \citenamefont
  {Braithwaite}, \citenamefont {Pourret}, \citenamefont {Aoki}, \citenamefont
  {Hardy}, \citenamefont {Flouquet},\ and\ \citenamefont
  {Brison}}]{Paulsen2020}%
  \BibitemOpen
  \bibfield  {author} {\bibinfo {author} {\bibfnamefont {C.}~\bibnamefont
  {Paulsen}}, \bibinfo {author} {\bibfnamefont {G.}~\bibnamefont {Knebel}},
  \bibinfo {author} {\bibfnamefont {G.}~\bibnamefont {Lapertot}}, \bibinfo
  {author} {\bibfnamefont {D.}~\bibnamefont {Braithwaite}}, \bibinfo {author}
  {\bibfnamefont {A.}~\bibnamefont {Pourret}}, \bibinfo {author} {\bibfnamefont
  {D.}~\bibnamefont {Aoki}}, \bibinfo {author} {\bibfnamefont {F.}~\bibnamefont
  {Hardy}}, \bibinfo {author} {\bibfnamefont {J.}~\bibnamefont {Flouquet}}, \
  and\ \bibinfo {author} {\bibfnamefont {J.~P.}\ \bibnamefont {Brison}},\
  }\href {http://arxiv.org/abs/2002.12724} {\bibfield  {journal} {\bibinfo
  {journal} {arXiv:2002.12724}\ } (\bibinfo {year} {2020})}\BibitemShut
  {NoStop}%
\bibitem [{\citenamefont {Schoenes}\ and\ \citenamefont
  {Franse}(1986)}]{Schoenes1986}%
  \BibitemOpen
  \bibfield  {author} {\bibinfo {author} {\bibfnamefont {J.}~\bibnamefont
  {Schoenes}}\ and\ \bibinfo {author} {\bibfnamefont {J.~J.~M.}\ \bibnamefont
  {Franse}},\ }\href
  {https://journals.aps.org/prb/pdf/10.1103/PhysRevB.33.5138} {\bibfield
  {journal} {\bibinfo  {journal} {Phys. Rev. B}\ }\textbf {\bibinfo {volume}
  {33}},\ \bibinfo {pages} {5138} (\bibinfo {year} {1986})}\BibitemShut
  {NoStop}%
\bibitem [{\citenamefont {Had{\v{z}}i{\'{c}}–Leroux}\ \emph
  {et~al.}(1986)\citenamefont {Had{\v{z}}i{\'{c}}–Leroux}, \citenamefont
  {Hamzi{\'{c}}}, \citenamefont {Fert}, \citenamefont {Haen}, \citenamefont
  {Lapierre},\ and\ \citenamefont {Laborde}}]{HadzicLeroux1986}%
  \BibitemOpen
  \bibfield  {author} {\bibinfo {author} {\bibfnamefont {M.}~\bibnamefont
  {Had{\v{z}}i{\'{c}}–Leroux}}, \bibinfo {author} {\bibfnamefont
  {A.}~\bibnamefont {Hamzi{\'{c}}}}, \bibinfo {author} {\bibfnamefont
  {A.}~\bibnamefont {Fert}}, \bibinfo {author} {\bibfnamefont {P.}~\bibnamefont
  {Haen}}, \bibinfo {author} {\bibfnamefont {F.}~\bibnamefont {Lapierre}}, \
  and\ \bibinfo {author} {\bibfnamefont {O.}~\bibnamefont {Laborde}},\ }\href
  {\doibase 10.1209/0295-5075/1/11/006} {\bibfield  {journal} {\bibinfo
  {journal} {Europhys. Lett.}\ }\textbf {\bibinfo {volume} {1}},\ \bibinfo
  {pages} {579} (\bibinfo {year} {1986})}\BibitemShut {NoStop}%
\bibitem [{\citenamefont {Yang}(2013)}]{Yang2013}%
  \BibitemOpen
  \bibfield  {author} {\bibinfo {author} {\bibfnamefont {Y.-F.}\ \bibnamefont
  {Yang}},\ }\href {\doibase 10.1103/PhysRevB.87.045102} {\bibfield  {journal}
  {\bibinfo  {journal} {Phys. Rev. B}\ }\textbf {\bibinfo {volume} {87}},\
  \bibinfo {pages} {45102} (\bibinfo {year} {2013})}\BibitemShut {NoStop}%
\bibitem [{\citenamefont {{Palacio Morales}}\ \emph {et~al.}(2015)\citenamefont
  {{Palacio Morales}}, \citenamefont {Pourret}, \citenamefont {Seyfarth},
  \citenamefont {Suzuki}, \citenamefont {Braithwaite}, \citenamefont {Knebel},
  \citenamefont {Aoki},\ and\ \citenamefont {Flouquet}}]{PalacioMorales2015}%
  \BibitemOpen
  \bibfield  {author} {\bibinfo {author} {\bibfnamefont {A.}~\bibnamefont
  {{Palacio Morales}}}, \bibinfo {author} {\bibfnamefont {A.}~\bibnamefont
  {Pourret}}, \bibinfo {author} {\bibfnamefont {G.}~\bibnamefont {Seyfarth}},
  \bibinfo {author} {\bibfnamefont {M.-T.}\ \bibnamefont {Suzuki}}, \bibinfo
  {author} {\bibfnamefont {D.}~\bibnamefont {Braithwaite}}, \bibinfo {author}
  {\bibfnamefont {G.}~\bibnamefont {Knebel}}, \bibinfo {author} {\bibfnamefont
  {D.}~\bibnamefont {Aoki}}, \ and\ \bibinfo {author} {\bibfnamefont
  {J.}~\bibnamefont {Flouquet}},\ }\href {\doibase 10.1103/PhysRevB.91.245129}
  {\bibfield  {journal} {\bibinfo  {journal} {Physical Review B}\ }\textbf
  {\bibinfo {volume} {91}},\ \bibinfo {pages} {245129} (\bibinfo {year}
  {2015})}\BibitemShut {NoStop}%
\bibitem [{\citenamefont {Fert}\ and\ \citenamefont {Levy}(1987)}]{Fert}%
  \BibitemOpen
  \bibfield  {author} {\bibinfo {author} {\bibfnamefont {A.}~\bibnamefont
  {Fert}}\ and\ \bibinfo {author} {\bibfnamefont {P.~M.}\ \bibnamefont
  {Levy}},\ }\href {https://journals.aps.org/prb/pdf/10.1103/PhysRevB.36.1907}
  {\bibfield  {journal} {\bibinfo  {journal} {Phys. Rev. B}\ }\textbf {\bibinfo
  {volume} {36}},\ \bibinfo {pages} {1907} (\bibinfo {year}
  {1987})}\BibitemShut {NoStop}%
\bibitem [{\citenamefont {Yamada}\ \emph {et~al.}(1993)\citenamefont {Yamada},
  \citenamefont {Ani}, \citenamefont {Kohno},\ and\ \citenamefont
  {Inagaki}}]{Yamada1993}%
  \BibitemOpen
  \bibfield  {author} {\bibinfo {author} {\bibfnamefont {K.}~\bibnamefont
  {Yamada}}, \bibinfo {author} {\bibfnamefont {K.}~\bibnamefont {Ani}},
  \bibinfo {author} {\bibfnamefont {H.}~\bibnamefont {Kohno}}, \ and\ \bibinfo
  {author} {\bibfnamefont {S.}~\bibnamefont {Inagaki}},\ }\href
  {https://www.researchgate.net/profile/Hiroshi{\_}Kohno/publication/311745768{\_}Anomalous{\_}Hall{\_}Coefficient{\_}in{\_}Heavy{\_}Electron{\_}Systems/links/599ceb10a6fdcc50034cb1f6/Anomalous-Hall-Coefficient-in-Heavy-Electron-Systems.pdf}
  {\bibfield  {journal} {\bibinfo  {journal} {Prog. Theor. Phys.}\ }\textbf
  {\bibinfo {volume} {89}} (\bibinfo {year} {1993})}\BibitemShut {NoStop}%
\bibitem [{\citenamefont {Palacio-Morales}\ \emph {et~al.}(2013)\citenamefont
  {Palacio-Morales}, \citenamefont {Pourret}, \citenamefont {Knebel},
  \citenamefont {Combier}, \citenamefont {Aoki}, \citenamefont {Harima},\ and\
  \citenamefont {Flouquet}}]{Palacio-Morales2013}%
  \BibitemOpen
  \bibfield  {author} {\bibinfo {author} {\bibfnamefont {A.}~\bibnamefont
  {Palacio-Morales}}, \bibinfo {author} {\bibfnamefont {A.}~\bibnamefont
  {Pourret}}, \bibinfo {author} {\bibfnamefont {G.}~\bibnamefont {Knebel}},
  \bibinfo {author} {\bibfnamefont {T.}~\bibnamefont {Combier}}, \bibinfo
  {author} {\bibfnamefont {D.}~\bibnamefont {Aoki}}, \bibinfo {author}
  {\bibfnamefont {H.}~\bibnamefont {Harima}}, \ and\ \bibinfo {author}
  {\bibfnamefont {J.}~\bibnamefont {Flouquet}},\ }\href {\doibase
  10.1103/PhysRevLett.110.116404} {\bibfield  {journal} {\bibinfo  {journal}
  {Phys. Rev. Lett.}\ }\textbf {\bibinfo {volume} {110}},\ \bibinfo {pages}
  {116404} (\bibinfo {year} {2013})}\BibitemShut {NoStop}%
\bibitem [{\citenamefont {Miao}\ \emph
  {et~al.}(2020{\natexlab{b}})\citenamefont {Miao}, \citenamefont {Liu},
  \citenamefont {Xu}, \citenamefont {Kotta}, \citenamefont {Kang},
  \citenamefont {Ran}, \citenamefont {Paglione}, \citenamefont {Kotliar},
  \citenamefont {Butch}, \citenamefont {Denlinger},\ and\ \citenamefont
  {Wray}}]{Miao2020}%
  \BibitemOpen
  \bibfield  {author} {\bibinfo {author} {\bibfnamefont {L.}~\bibnamefont
  {Miao}}, \bibinfo {author} {\bibfnamefont {S.}~\bibnamefont {Liu}}, \bibinfo
  {author} {\bibfnamefont {Y.}~\bibnamefont {Xu}}, \bibinfo {author}
  {\bibfnamefont {E.~C.}\ \bibnamefont {Kotta}}, \bibinfo {author}
  {\bibfnamefont {C.-J.}\ \bibnamefont {Kang}}, \bibinfo {author}
  {\bibfnamefont {S.}~\bibnamefont {Ran}}, \bibinfo {author} {\bibfnamefont
  {J.}~\bibnamefont {Paglione}}, \bibinfo {author} {\bibfnamefont
  {G.}~\bibnamefont {Kotliar}}, \bibinfo {author} {\bibfnamefont {N.~P.}\
  \bibnamefont {Butch}}, \bibinfo {author} {\bibfnamefont {J.~D.}\ \bibnamefont
  {Denlinger}}, \ and\ \bibinfo {author} {\bibfnamefont {L.~A.}\ \bibnamefont
  {Wray}},\ }\href {\doibase 10.1103/PhysRevLett.124.076401} {\bibfield
  {journal} {\bibinfo  {journal} {Phys. Rev. Lett.}\ }\textbf {\bibinfo
  {volume} {124}},\ \bibinfo {pages} {076401} (\bibinfo {year}
  {2020}{\natexlab{b}})}\BibitemShut {NoStop}%
\end{thebibliography}
%

\end{document}


\title{\textbf{Supplemental Material: Evidence of  Fermi Surface Reconstruction at the Metamagnetic Transition of the Strongly Correlated Superconductor UTe$_2$}}

\author{Q.~Niu}
\affiliation{Univ. Grenoble Alpes, CEA, IRIG, PHELIQS, F-38000 Grenoble, France}
\author{G.~Knebel}
\affiliation{Univ. Grenoble Alpes, CEA, IRIG, PHELIQS, F-38000 Grenoble, France}
\author{D.~Braithwaite }
\affiliation{Univ. Grenoble Alpes, CEA, IRIG, PHELIQS, F-38000 Grenoble, France}
\author{D.~Aoki }
\affiliation{Univ. Grenoble Alpes, CEA, IRIG, PHELIQS, F-38000 Grenoble, France}
\affiliation{Institute for Materials Research, Tohoku University, Oarai, Ibaraki, 311-1313, Japan}
\author{G.~Lapertot}
\affiliation{Univ. Grenoble Alpes, CEA, IRIG, PHELIQS, F-38000 Grenoble, France}
\author{M.~Vali\v{s}ka}
\affiliation{Univ. Grenoble Alpes, CEA, IRIG, PHELIQS, F-38000 Grenoble, France}
\author{G.~Seyfarth}
\affiliation{Univ. Grenoble Alpes, EMFL, CNRS, Laboratoire National des Champs Magn\'etiques Intenses (LNCMI), 38042 Grenoble, France}
\author{W.~Knafo}
\affiliation{Laboratoire National des Champs Magnétiques Intenses, UPR 3228, CNRS-UPS-INSA-UGA,143 Avenue de Rangueil, 31400 Toulouse, France}
\author{T. Helm}
\affiliation{Max Planck Institute for Chemical Physics of Solids, 01187 Dresden, Germany}
\affiliation{Dresden High Magnetic Field Laboratory (HLD-EMFL), Helmholtz-Zentrum Dresden-Rossendorf, 01328 Dresden, Germany}
\author{J-P.~Brison}
\affiliation{Univ. Grenoble Alpes, CEA, IRIG, PHELIQS, F-38000 Grenoble, France}
\author{J.~Flouquet}
\affiliation{Univ. Grenoble Alpes, CEA, IRIG, PHELIQS, F-38000 Grenoble, France}
\author{A.~Pourret}
\affiliation{Univ. Grenoble Alpes, CEA, IRIG, PHELIQS, F-38000 Grenoble, France}
\email[E-mail me at: ]{alexandre.pourret@cea.fr}
\date{\today } 


\maketitle

\subsection*{I. Hysteresis in $S(H)$ of $\mathrm{UTe}_2$ at $H_\mathrm{m}$}

\begin{figure}[h]
              \includegraphics[width=0.7\textwidth]{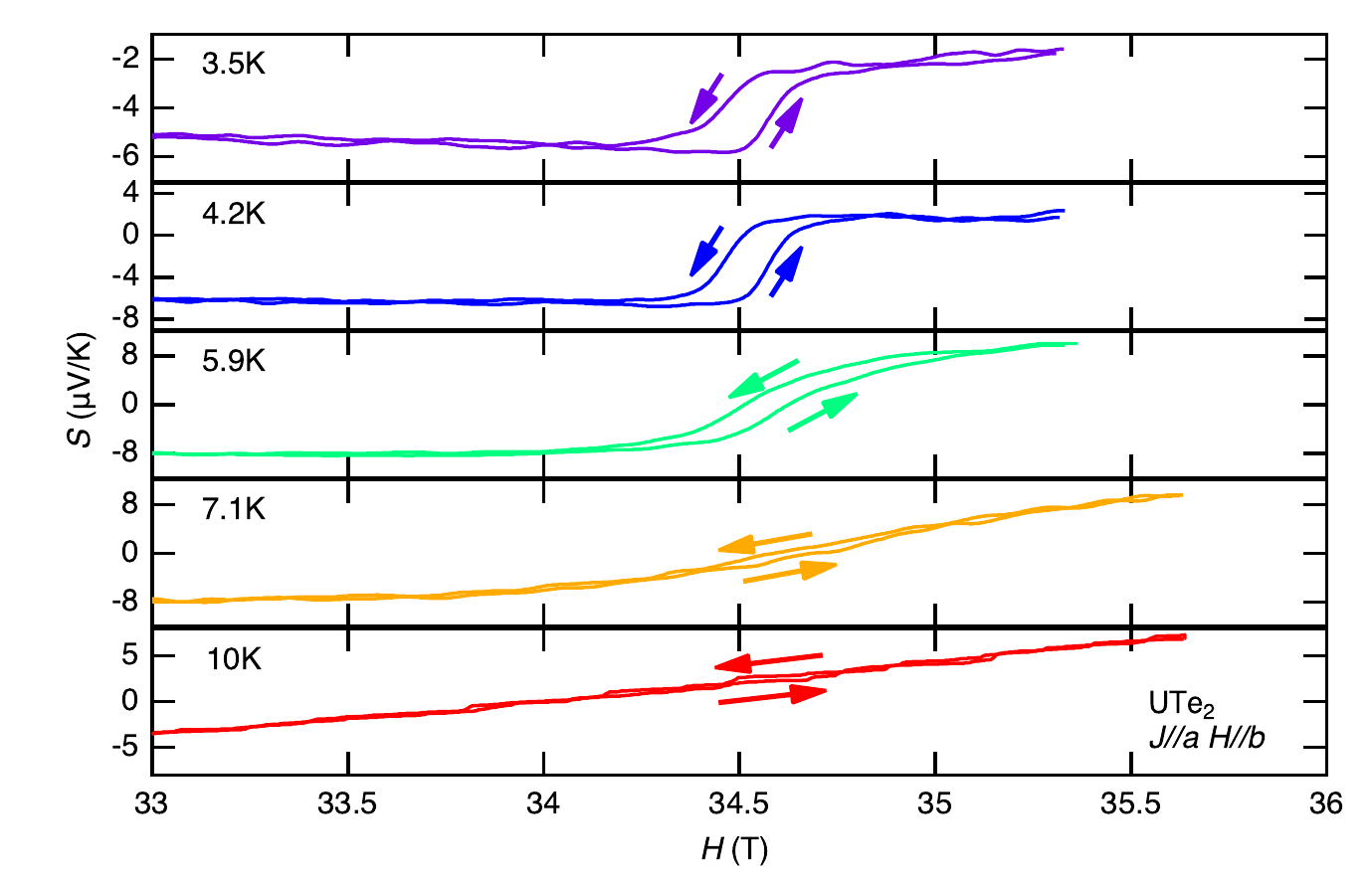}    
              \caption{\label{FigS1} (color online) Temperature evolution of the hysteresis observed in $S$ at $H_\mathrm{m}$. The arrows indicate the direction of the field sweep. The first-order transition ends at the critical end point $T_\mathrm{CEP}\approx$7~K.}
\end{figure}

With the same configuration as described in the main text, we measured $S(H)$ in a narrow field range near $H_\mathrm{m}$ sweeping the field up and down  through $H_m$ with lower ramping rate. Fig.~\ref{FigS1} shows the results at different temperatures up to 10~K. A hysteresis is clearly observed at low temperatures revealing the first-order character of the metamagnetic transition (MMT). The vanishing of the hysteresis with increasing temperature indicates that the first-order transition terminates at the critical end point (CEP) with $T_\mathrm{CEP}\approx$7~K.

\subsection*{II. Resistivity and Hall signal for $H \parallel b$}

\begin{figure}[h]
             \includegraphics[width=0.8\textwidth]{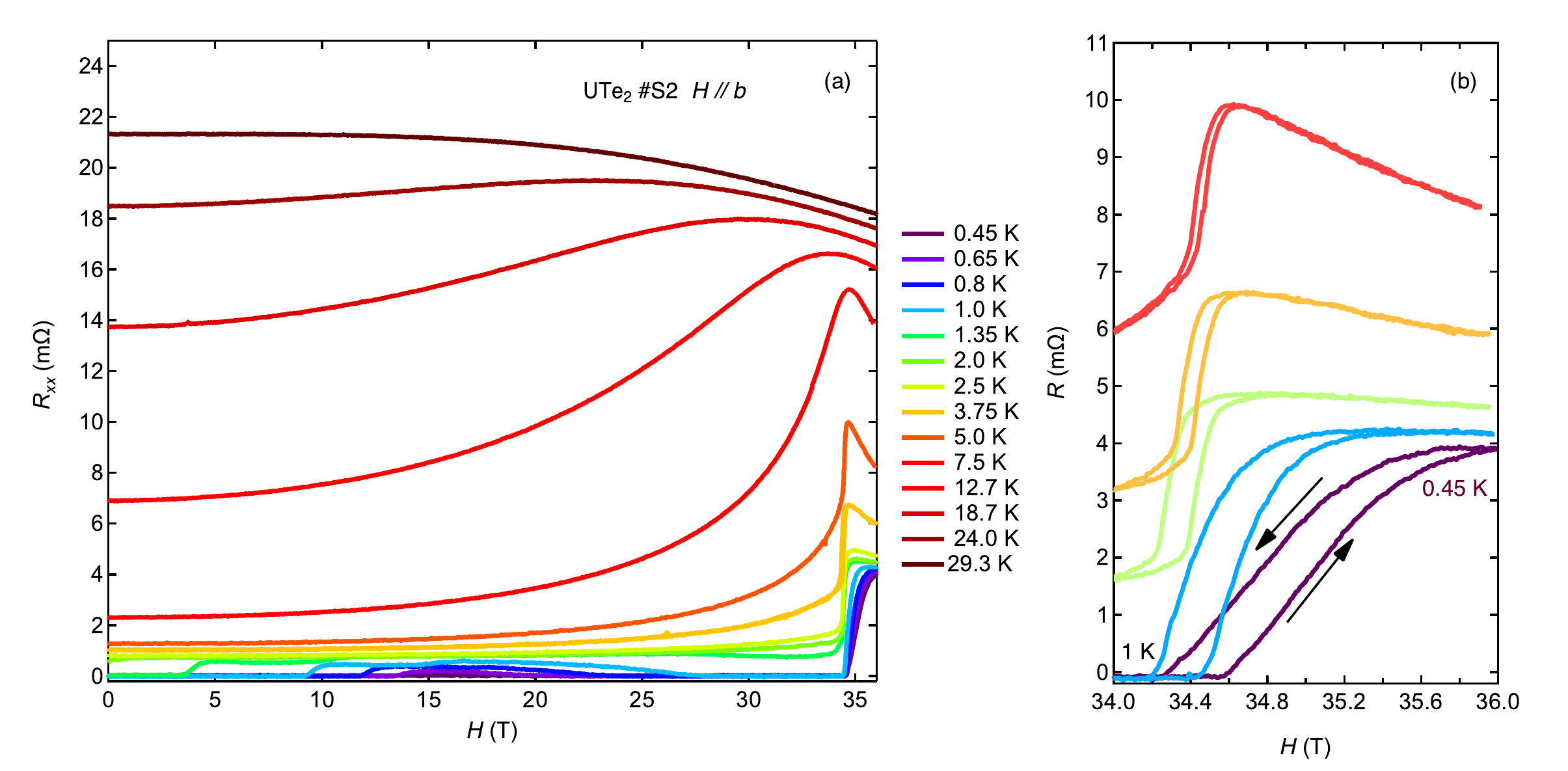}      
              \caption{\label{FigS2} (color online) (a) Field dependence of $R_{xx}$ on sample S2 up to 36~T at different temperatures for $H \parallel b$. (b) Hysteresis loop of $R_{xx}$ at the metamagnetic transition highlighting the first order character of the transition below $T_\mathrm{CEP}\approx7$~K.}
\end{figure}

\begin{figure}[h]
              \includegraphics[width=0.8\textwidth]{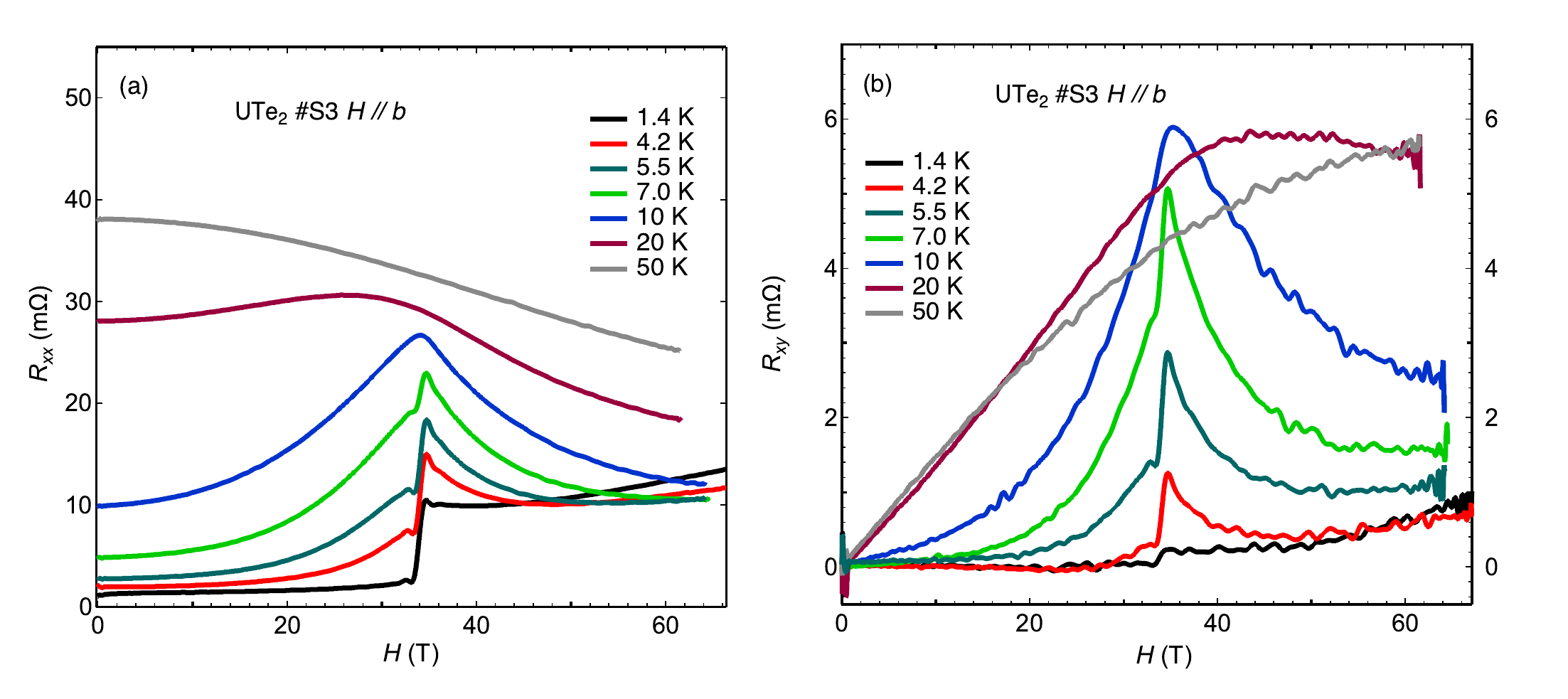}     
              \caption{\label{FigS3} (color online) Field dependence of $R_{xx}$ (a) and $R_{xy}$ (b) on sample S3 up to 68~T at different temperatures.}
\end{figure}

\begin{figure}[h]
              \includegraphics[width=0.7\textwidth]{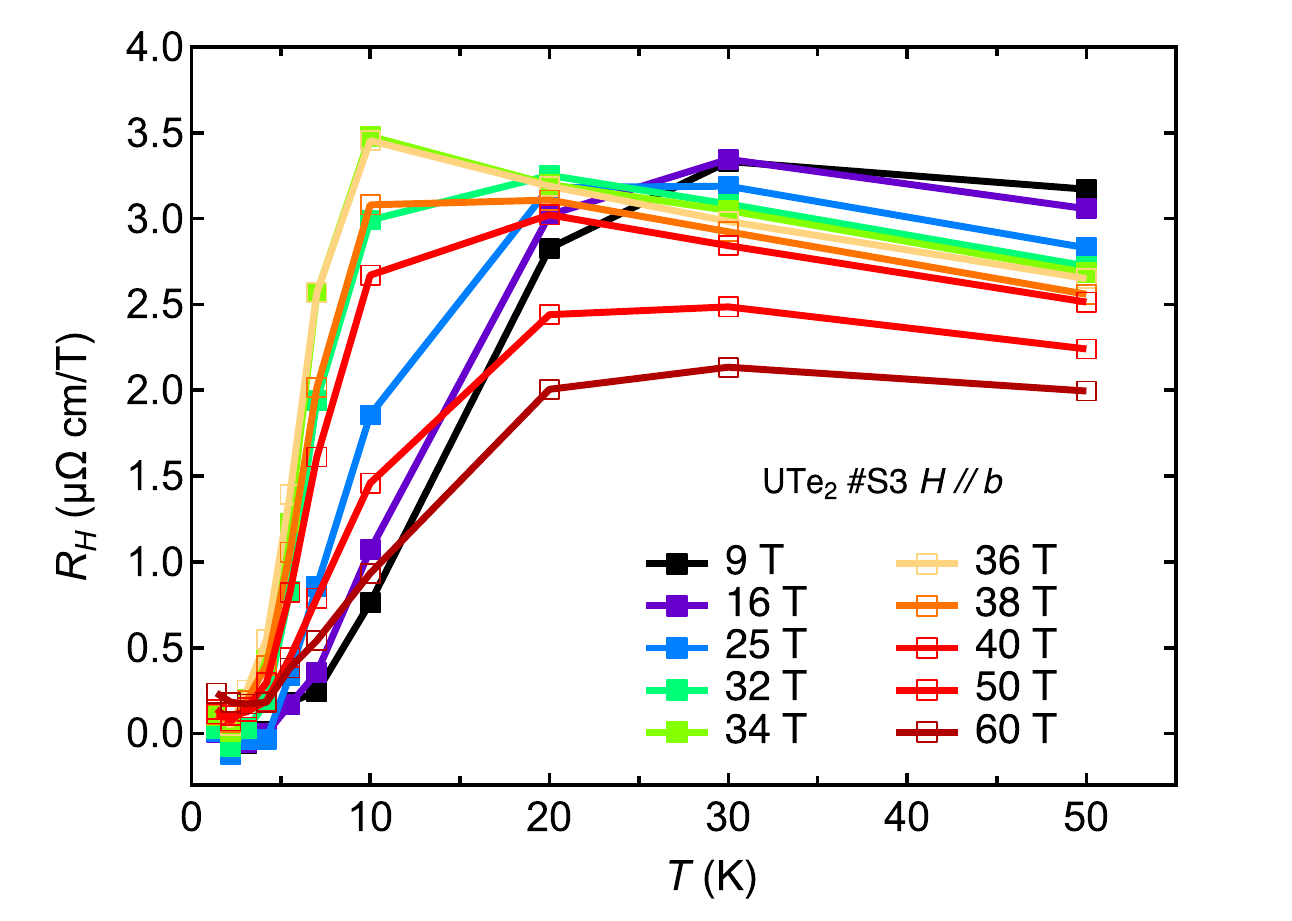}        
              \caption{\label{FigS4} (color online) Temperature dependence of $R_\mathrm{H}$ on sample S3 at different magnetic fields for $H<H_\mathrm{m}$ (full symbols) and for $H>H_\mathrm{m}$ (open symbols).}
\end{figure}

Resistivity and Hall effect have been measured on sample S2 and S3, part of the results have been already shown in the main text. Fig.~\ref{FigS2}(a) shows the magnetic field dependence of  $R_{xx}$ for different temperatures up to 36~T for sample S2. Below 1.5~K, $R_{xx}$ is zero up to $H_\mathrm{m}$ and then shows a step like anomaly indicating that $H_\mathrm{C2}$ is equal to $H_\mathrm{m}$ for  $H \parallel b$. By increasing temperature, $R_{xx}$ is non zero below $H_m$ but still shows a step like anomaly at the MMT up to $T_\mathrm{CEP}\approx7$~K. Above $T_\mathrm{CEP}$, the transition broadens significantly. The first order character of the transition below $T_\mathrm{CEP}$ is also confirmed by the observation of an hysteresis loop in the resistivity, see Fig.~\ref{FigS2}(b). Interestingly, as soon as the system becomes superconducting below 1.5~K, the width of the transition broadens and the superconducting transition seems to move to higher field, conserving a hysteresis loop. 
The field dependence of $R_{xx}$ and $R_{xy}$ up to 68~T of sample S3 are represented in Fig.~\ref{FigS3}. The resistivity data are similar to sample S2. At low temperature, $R_{xy}$ is almost field independent below $H_m$, shows a small step like anomaly at $H_m$ and becomes field dependent above $H_\mathrm{m}$. By increasing  temperature, the transition becomes a peak which is very sharp at $T_\mathrm{CEP}\approx7$~K and broadens for high temperature.
Fig.~\ref{FigS4} shows the temperature dependence of the Hall coefficient $R_\mathrm{H}$ measured on sample S3. From low magnetic field, the temperature of the maximum in $R_\mathrm{H}(T)$, defined as $T_\mathrm{m}$ shifts to low temperature when approaching $H_\mathrm{m}$ down to $T_\mathrm{CEP}$. Above $H_m$,  the temperature of the maximum $T_\mathrm{cr}$, corresponding to the crossover temperature between the PM and PPM state, shifts to higher temperature. The energy scale associated to $T_\mathrm{m}$ (similar to  $T_{\chi_\mathrm{max}}$) doesn't decrease to zero temperature but is limited to 
$T_\mathrm{CEP}$ due to the first order nature of the transition and so to the absence of a quantum critical point.

\subsection*{III. Ordinary and anomalous Hall effect}
\begin{figure}[h!]
              \includegraphics[width=0.7\textwidth]{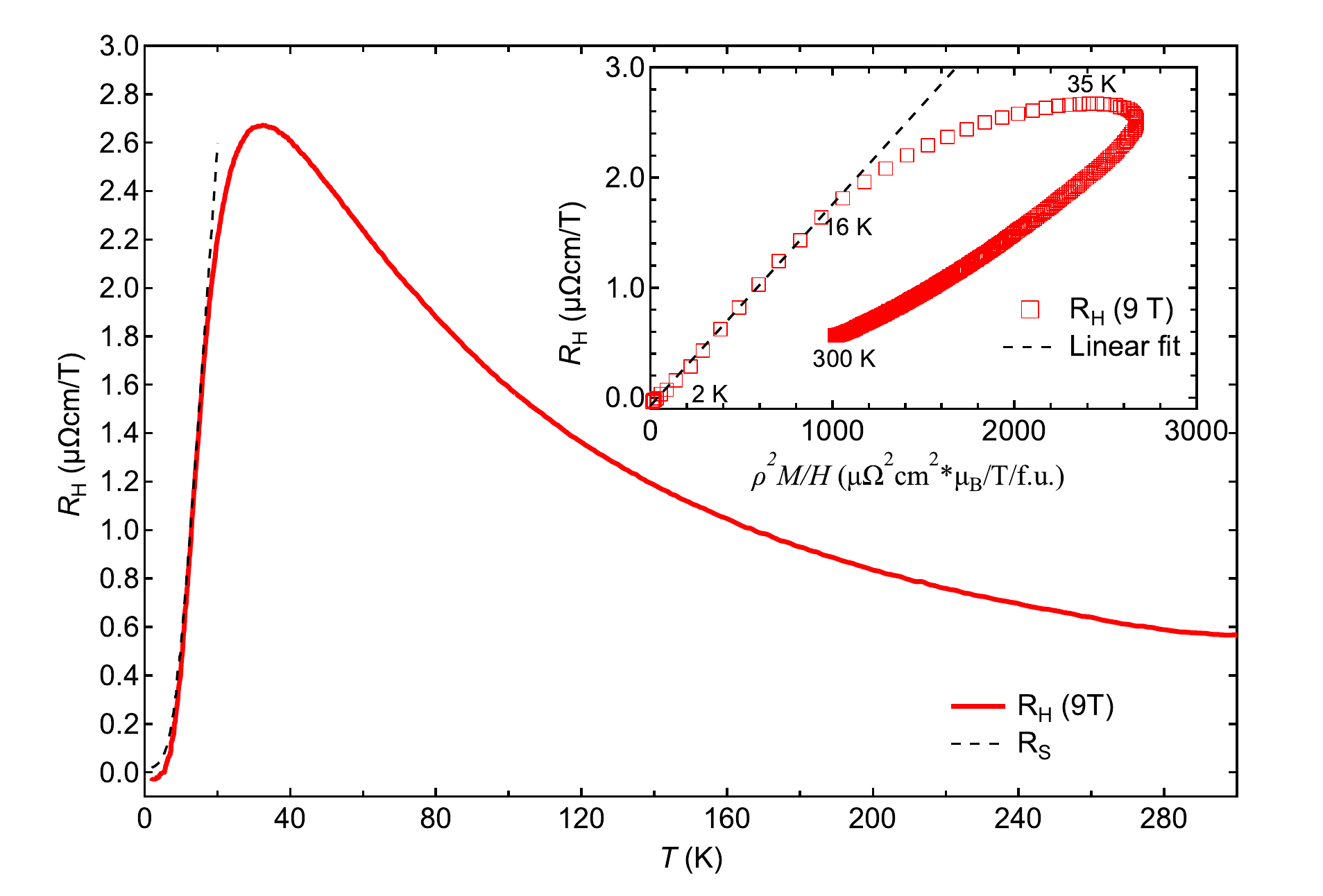}        
              \caption{\label{FigS4b} (color online) Temperature dependence of $R_\mathrm{H}$ between 1.4~K and 300~K. The different fitting curves for the anomalous Hall effect at low temperature ($\propto  \rho^2 M/H$) and high temperature ($ \propto \rho M/H$) are represented (dashed lines). Inset:  $R_\mathrm{H}$  as a function of  $\rho^2 M/H$. The dash line represents linear fitting.}
\end{figure}

Here we describe the procedure to estimate the ordinary and anomalous contribution to the Hall effect. The first assumption is that the Hall signal is the sum of two terms, $R_\mathrm{H}=R_0+R_\mathrm{S}$ where $R_0$ is the ordinary part associated to de density of carriers and their mobility and $R_\mathrm{S}$ is the anomalous part associated to scattering processes on magnetic impurities. The microscopic origin of $R_\mathrm{S}$ is quite complex. Three distinct contributions, intrinsic, skew scattering, and side jump scattering have been identified. Each of them has an individual scaling $R_\mathrm{S} \propto \rho^{\alpha} M_\mathrm{z}/H$ with respect to the longitudinal resistivity $\rho$. Here, $M_\mathrm{z}$ is the magnetization and $H$ the magnetic field along $z$ axis. In ferromagnetic materials, the summation of the three terms yields an empirical formula that explains a large amount of experimental data. However, in heavy fermion materials a satisfactory formula has not been achieved nevertheless it have been observed that the skew scattering is the dominant scattering process with two different scaling depending of the temperature\cite{Fert, Yang2013}. At high temperature, for $T>T^*$ (coherence temperature), the incoherent skew scattering of conduction electrons by independent f electrons should be considered and then the relation $R_\mathrm{S}=C'\times \rho M/H$ is expected. On the other hand, at low temperature for $T<T_{FL}$ (Fermi liquid temperature) a different scaling is expected, $R_\mathrm{S}=C \times \rho^2 M/H$ , due to the coherent skew scattering of  f electrons once the Fermi surface is well defined. Between $T^*$ and $T_{FL}$ a cross-over regime is observed. In Fig.~\ref{FigS4b}, the temperature dependence of $R_\mathrm{H}$ is represented. The inset shows that as expected at low temperature (for $T< 20$~K), the Hall signal is well described by a linear fitting as a function of $\rho^2 M/H$ indicating that below $T_{\chi_\mathrm{max}}$, a coherent regime appears. The fact that the fit ($R_\mathrm{S}\_LT$) is smaller than the $R_\mathrm{H}$ data at very low temperatures indicates that the anomalous contribution to the Hall signal is negligible at low temperature, and justifies to extract the number of carriers by using the Hall signal above $T_\mathrm{SC}$ in this material \cite{Niu2019}. The fit using $\rho M/H$, ($R_\mathrm{S}\_HT$), is not very satisfactory at high temperature. This discrepancy can be explained by the phonon contribution in the resistivity \cite{Fert}.

\begin{figure}
              \includegraphics[width=0.6\textwidth]{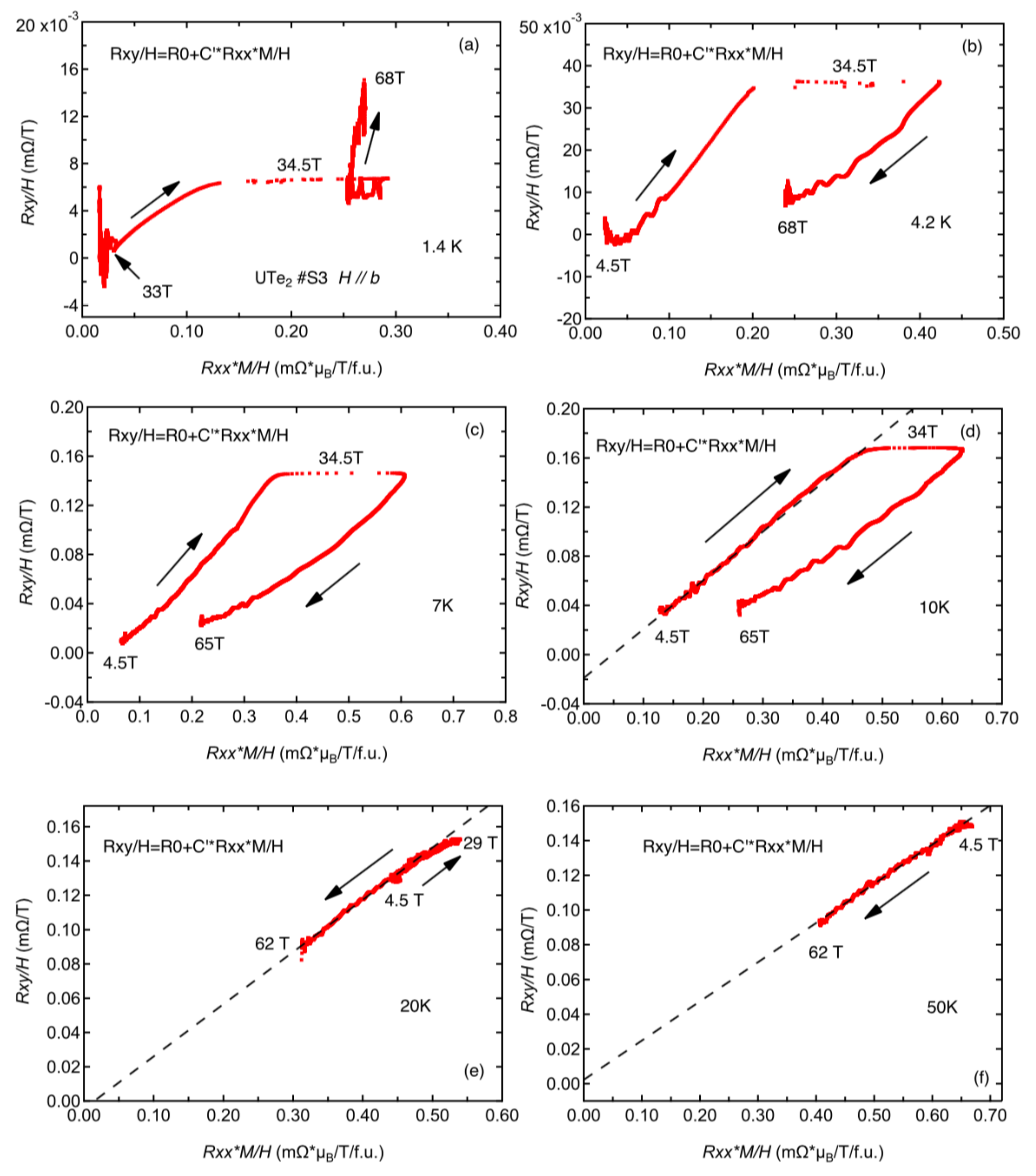}        
             \caption{\label{FigS5} (color online) $R_{xy}/H$ against $R_{xx}M/H$. The dashed lines are linear fits.}
              \includegraphics[width=0.6\textwidth]{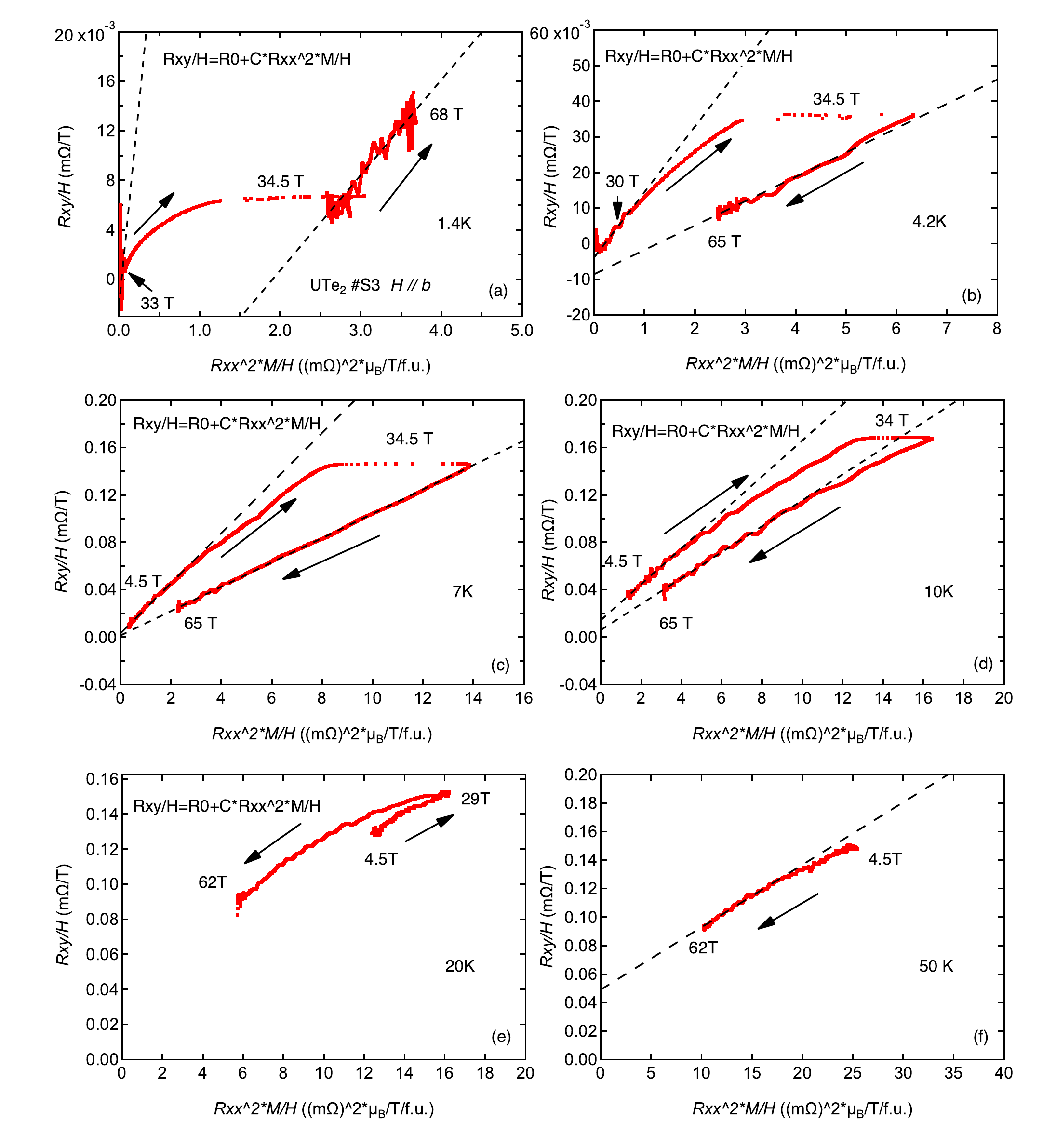}        
              \caption{\label{FigS6} (color online) $R_{xy}/H$ against $R_{xx}^2M/H$. The dashed lines are linear fits.}
\end{figure}
\begin{figure}
              \includegraphics[width=0.6\textwidth]{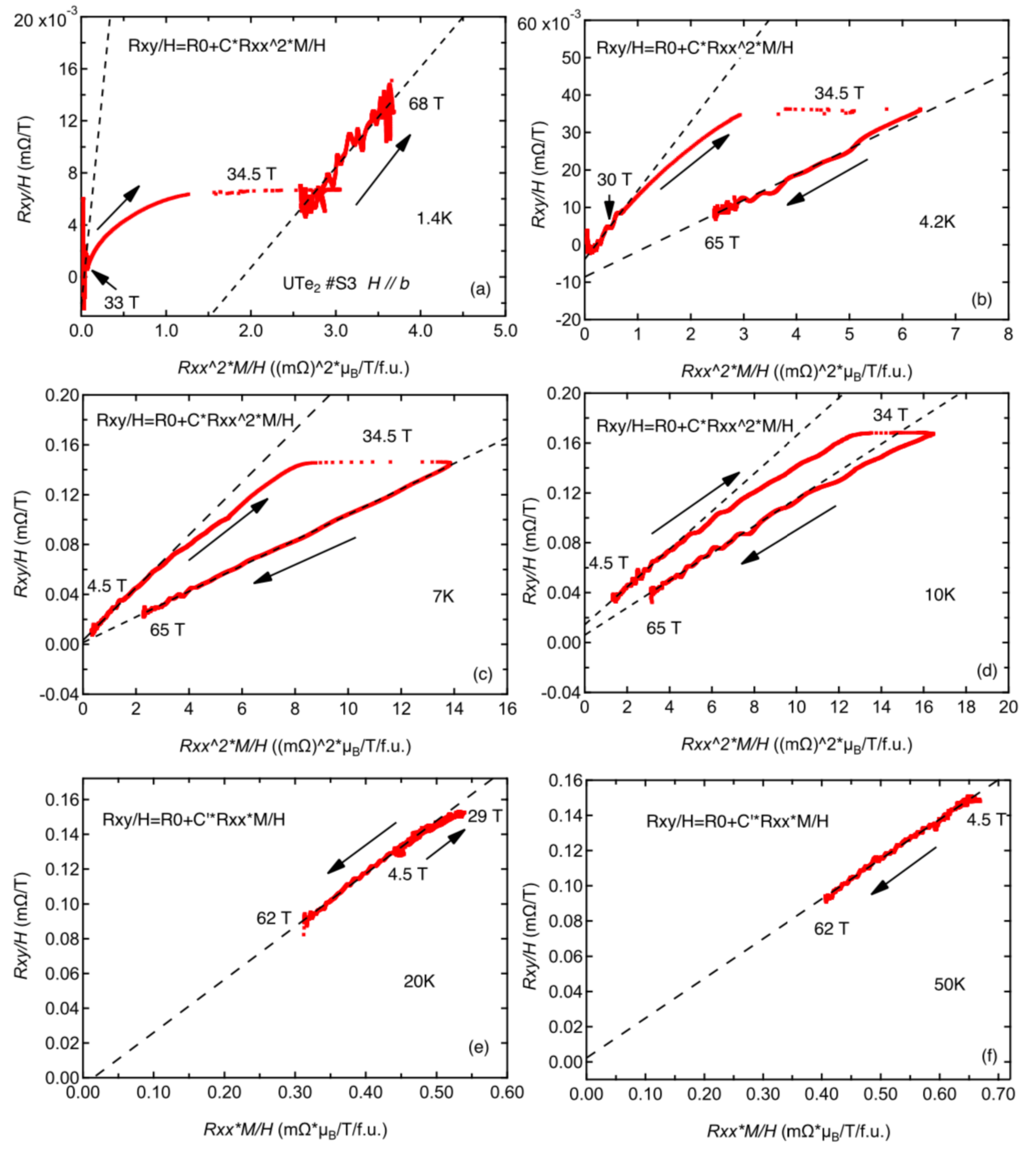}         
              \caption{\label{FigS7} (color online) $R_{xy}/H$ against $R_{xx}^2M/H$ at 1.4~K, 4.2~K, 7~K and 10~K. $R_{xy}/H$ against $R_{xx}M/H$ at 20~K and 50~K. The dashed lines are linear fits.}
\end{figure}

To extract the coefficients $C$ and $C'$, we have plotted $R_{xy}/H$ as a function of  $R_{xx} M/H$ in figure Fig.~\ref{FigS5} and as a function of  $R_{xx}^2 M/H$ in Fig.~\ref{FigS6}. We observe that for  $T<20~K$, the data are well described by  $R_{xx}^2 M/H$ meaning that below $T^*$ (roughly $T \chi_{max}$), the  anomalous Hall signal is well described by coherent skew scattering of f electrons. For high temperature, the anomalous Hall signal is well described by $R_{xx} M/H$ underlying the predominance of incoherent skew scattering processes above $T^*$. These observations are summarised in Fig.~\ref{FigS7} where $R_{xy}/H$ is plotted either as a function of $R_{xx} M/H$ or $R_{xx}^2 M/H$ depending on the temperature. It is interesting to notice that for $T >20K$, for the whole magnetic field range, the data are well fitted by an unique linear function of  $R_{xx} M/H$ passing through the origin meaning that the ordinary Hall effect is negligible at high temperature. At low temperature, the data are clearly well described by two linear functions of  $R_{xx}^2 M/H$ below and above $H_m$. The value of the coefficients $C$ and $C'$ are summarised in table S1. The variation of $C$ indicates that the amplitude of the scattering changes through $H_\mathrm{m}$. At 1.4~K, below $H_m$ the data don't show any linear dependence meaning that the Hall signal is ordinary. On the other hand, above $H_m$, the  data show still linear dependence meaning that even at very low temperature, there is still a contribution from the anomalous Hall effect above the metamagnetic transition just above $T_\mathrm{SC}$.

\begin{table*}
\begin{ruledtabular}
\caption{\label{Table}List of anomalous Hall effect coefficients $C$ ($R_\mathrm{S}=C \times \rho^2 M/H$) for $H$ below and above $H_m$ and  $C'$ ($R_\mathrm{S}=C' \times \rho M/H$) .}
\begin{tabular}{|c|cc|c|}  
  \multicolumn{1}{|c|} { T (K) }&  \multicolumn{2}{c|}{$C$(arb. units)} & $C(H>H_m)/C(H<H_m$) \\

 \multicolumn{1}{|c|}{} & \multicolumn{1}{c} {$H<H_m$} & \multicolumn{1}{c|} {$H>H_m$}  &  \multicolumn{1}{c|}{}\\
  1.4 & 65.3 & 8.6 & 0.13  \\   
  4.2 & 18.4 & 7.6 & 0.41 \\
  7    & 21.1   & 10.2 &0.48 \\ 
  10 & 13.9   & 10.5 & 0.75  \\
     \hline
    \multicolumn{1}{|c|} { }&  \multicolumn{2}{c|}{$C'$(arb. units)} &  \\
   20 & \multicolumn{2}{c|}{0.30}  &  \\
   50  & \multicolumn{2}{c|}{0.22}  &  \\
  
\end{tabular}
\end{ruledtabular}
\end{table*}

\begin{figure}
              \includegraphics[width=0.7\textwidth]{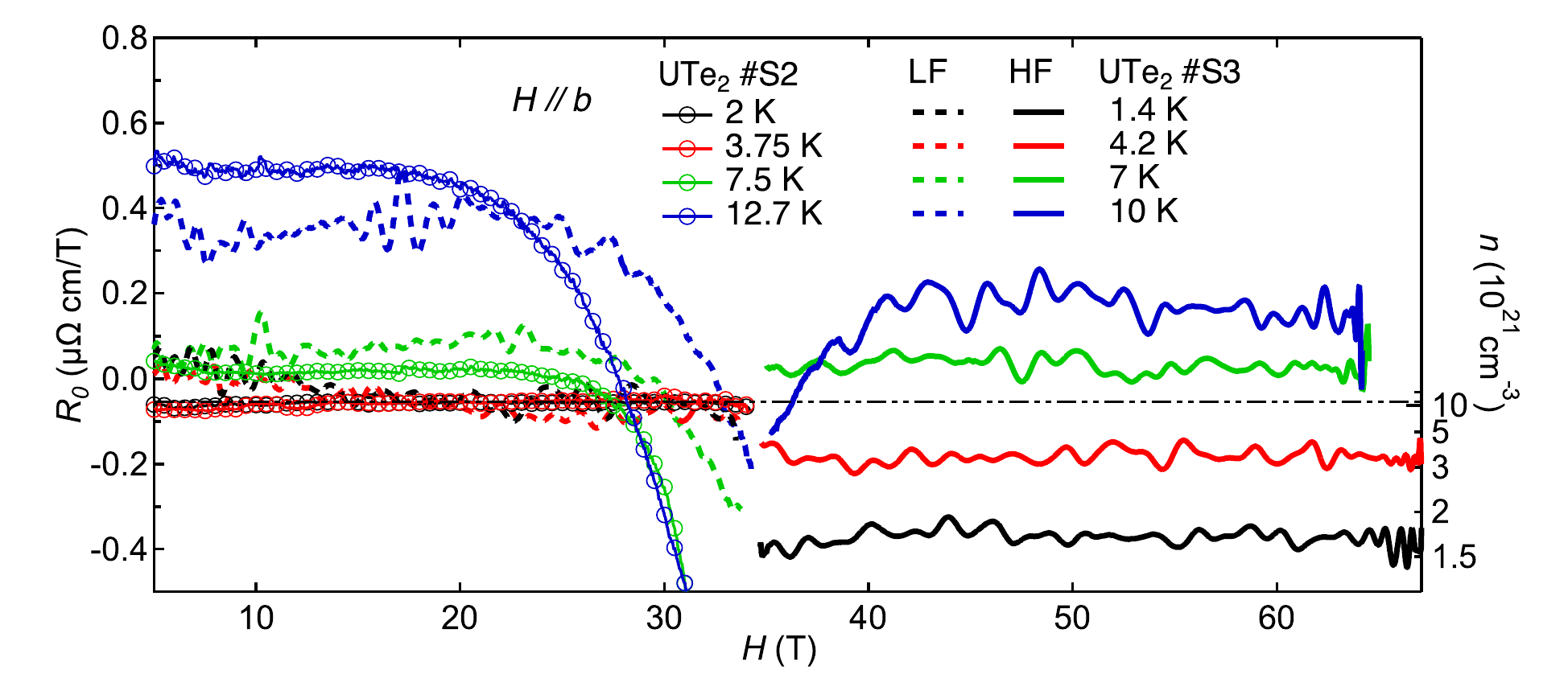}        
              \caption{\label{FigS9} (color online) Complement to Fig. 4 b) of main text. Ordinary Hall effect as a function of magnetic field for different temperatures obtained after subtracting the anomalous contribution (taking into account its change of amplitude hrough $H_m$). This implies two sets of data for low field ( dashed lines LF) and high field ( full lines HF) for the sample S3 measured up to 68 T. The right scale indicates the carrier density and the dashed-dot line represents the value obtained previously \cite{Niu2019}.}
              
\end{figure}

After extracting the contribution of the anomalous Hall effect, through the different coefficients $C$ and $C'$ depending on the temperature, we obtained the ordinary contribution by subtracting the anomalous part from the raw data by taking into account the change of the coefficient $C$ through $H_m$. For all temperatures, the field dependence of the ordinary  Hall effect is represented in Fig.~\ref{FigS9}. A drastic change of the ordinary Hall effect at $H_m$ is extracted from this analysis indicating a change of the number of carriers associated or not with a change of their mobility.

\bibliographystyle{apsrev4-1}